\documentclass[fleqn,usenatbib]{mnras}

\usepackage{newtxtext,newtxmath}

\usepackage[T1]{fontenc}

\DeclareRobustCommand{\VAN}[3]{#2}
\let\VANthebibliography\thebibliography
\def\thebibliography{\DeclareRobustCommand{\VAN}[3]{##3}\VANthebibliography}

\usepackage{graphicx}	
\usepackage{amsmath}	
\usepackage{gensymb}    
\usepackage{xspace}
\usepackage{multirow}
\usepackage{pifont}
\usepackage{mathtools}  
\usepackage{bigdelim, makecell, booktabs}  

\newcommand{\msol}{${\rm M_{\odot}}$\xspace}
\newcommand{\costuum}{{\sc CosTuuM}\xspace}

\newcommand{\fac}{$f_{\rm aC}$\xspace}

\newcommand{\tac}{$T_{\rm aC}$\xspace}
\newcommand{\tsil}{$T_{\rm Sil}$\xspace}

\title[Crab HAWC+ polarisation]{SOFIA/HAWC+ observations of the Crab Nebula: dust properties from polarised emission}

\author[J. Chastenet, I. De Looze et al.]{
J\'er\'emy Chastenet,$^{1}$\thanks{jeremy.chastenet@ugent.be}
Ilse De Looze,$^{1,2}$\thanks{ilse.delooze@ugent.be}
Brandon S. Hensley,$^{3}$
Bert Vandenbroucke,$^{4}$
\newauthor 
Mike J. Barlow,$^{2}$
Jeonghee Rho,$^{5}$
Aravind P. Ravi,$^{6}$
Haley L. Gomez,$^{7}$
\newauthor 
Florian Kirchschlager,$^{1,2}$
Juan Mac\'ias-P\'erez,$^8$
Mikako Matsuura,$^{7}$
Kate Pattle,$^{2}$
\newauthor 
Nicolas Ponthieu,$^{10}$
Felix D. Priestley,$^{7}$
Monica Relaño,$^{11,12}$
Alessia Ritacco, $^{13,14}$
Roger Wesson$^{2}$\\
$^{1}$Sterrenkundig Observatorium, Ghent University, Krijgslaan 281-S9, B9000 Gent, Belgium\\
$^{2}$Dept. of Physics \& Astronomy, University College London, Gower Street, London WC1E 6BT, UK\\
$^{3}$Department of Astrophysical Sciences,  Princeton University, Princeton, NJ 08544, USA\\
$^{4}$Leiden Observatory, Leiden University, NL-2300 RA Leiden, Netherlands\\
$^{5}$SETI Institute, 189 N. Bernardo Ave., Ste. 200, Mountain View, CA 94043\\
$^{6}$Box 19059, Department of Physics, University of Texas at Arlington, Arlington, TX 76019, USA\\
$^{7}$School of Physics and Astronomy, Cardiff University, Queens Buildings, The Parade, Cardiff CF24 3AA \\
$^{8}$Univ. Grenoble Alpes, CNRS, Grenoble INP, LPSC-IN2P3, 53, avenue des Martyrs, 38000 Grenoble, France\\
$^9$Centre for Astronomy, School of Physics, National University of Ireland Galway, University Road, Galway H91 TK33, Ireland\\
$^{10}$Univ. Grenoble Alpes, CNRS, IPAG, 38000 Grenoble, France\\
$^{11}$Departamento F\'isica Te\'orica y del Cosmos, Universidad de Granada, E-18071 Granada, Spain\\
$^{12}$Instituto Universitario Carlos I de F\'isica Te\'orica y Computacional, Universidad de Granada, E-18071 Granada, Spain\\
$^{13}$Institut d'Astrophysique Spatiale (IAS), CNRS, Universit\'e Paris Sud, Orsay, France\\
$^{14}$Laboratoire de Physique de l'\'Ecole Normale Sup\'erieure, ENS, PSL Research University, CNRS, Sorbonne Universit\'e, Universit\'e de Paris, 75005, Paris, France
}

\date{Accepted XXX. Received YYY; in original form ZZZ}

\pubyear{2022}

\begin{document}
\label{firstpage}
\pagerange{\pageref{firstpage}--\pageref{lastpage}}
\maketitle

\begin{abstract}
Supernova remnants (SNRs) are well-recognised dust producers, but their net dust production rate remains elusive due to uncertainties in grain properties that propagate into observed dust mass uncertainties, and determine how efficiently these grains are processed by reverse shocks. In this paper, we present a detection of polarised dust emission in the Crab pulsar wind nebula, the second SNR with confirmed polarised dust emission after Cassiopeia~A. We constrain the bulk composition of the dust with new SOFIA/HAWC+ polarimetric data in band C 89~$\mu$m and band D 154~$\mu$m. After correcting for synchrotron polarisation, we report dust polarisation fractions ranging between $3.7-9.6$~per~cent and $2.7-7.6$~per~cent in three individual dusty filaments at 89 and 154~$\mu$m, respectively. The detected polarised signal suggests the presence of large ($\gtrsim 0.05-0.1~\mu$m) grains in the Crab Nebula. With the observed polarisation, and polarised and total fluxes, we constrain the temperatures and masses of carbonaceous and silicate grains. We find that the carbon-rich grain mass fraction varies between 12 and 70~per~cent, demonstrating that carbonaceous and silicate grains co-exist in this SNR. Temperatures range from $\sim 40~$K to $\sim 70~$K and from $\sim 30~$K to $\sim 50~$K for carbonaceous and silicate grains, respectively. Dust masses range from $\sim 10^{-4}~$\msol to $\sim 10^{-2}~$\msol for carbonaceous grains and to $\sim 10^{-1}~$\msol for silicate grains, in three individual regions.
\end{abstract}

\begin{keywords}
ISM: supernova remnants -- (stars:) supernovae: individual: Crab -- (ISM:) dust, extinction -- polarization
\end{keywords}


\section{Introduction}
The total dust budget of a galaxy is the result of a complex network of dust grain formation and destruction processes. 
Evolved stars on the Asymptotic Giant Branch (AGB) form dust grains in their atmospheres, rich in metals from the nucleosynthesis in their cores \citep[e.g.][]{Gail1985, Gail1998FaDi, Gail1999, Ferrarotti2006, DellAgli2017, Hofner2018, Bladh2019, Nanni2019, VandeSande2021}. 
Supernovae (SNe) and supernova remnants (SNRs) are also believed to form grains in the ejecta where metals formed in the progenitors are released during their subsequent explosion \citep{Barlow2010, Matsuura2015, DeLooze2017, Cigan2019, Chawner2019, Chawner2020}. 
But the strong shocks and winds from that explosion, hot (SN-)gas, and cosmic rays may also lead to the fragmentation and/or destruction of pre-existing dust \citep{Jones1994, Jones1996, Nozawa2006, Bocchio2016, Slavin2015, Kirchschlager2022} or newly formed grains \citep{Kirchschlager2019, Slavin2020}. 
It is thus almost surprising that dust in SNRs is robustly detected often in large amounts.

To balance efficient dust destruction processes, dust growth in diffuse and/or dense parts of the interstellar medium (ISM) has been suggested to play a major role in the total dust mass budget \citep{Draine2009, Schneider2016, Zhukovska2016, DeVis2017, Zhukovska2018, Vilchez2019, DeVis2021, Galliano2021}. However, several works have demonstrated severe shortcomings in our current understanding of interstellar grain growth processes \citep[e.g.,][]{Ferrara2016, Ceccarelli2018, Priestley2021b}. For example, they show that the low density of the diffuse ISM, or the high dust temperatures of high-$z$ galaxies can hamper grain growth, or that the conditions for Coulomb attraction are not easily met.
The intricate inter-relations of ISM processes make it difficult to draw a detailed account of the processes that increase or decrease the dust content of a galaxy. Recent work has suggested that present-day dust budgets could be accounted for by stellar dust production if the efficiency of supernova shocks to destroy interstellar dust has been overestimated \citep{Matsuura2009, Jones&Nuth2011, DeLooze2020, Ferrara2021, Priestley2021a} and/or the effect of galactic outflows has been improperly accounted for \citep{Nanni2020}.
There are also hints that SN-formed silicate could be more abundant than previously initially thought \citep[][]{Leitner2019}.
Additionally, the early and recent Universe dust sources are likely different. In high-$z$ galaxies, AGB stars could not contribute significantly to the dust budget \citep[e.g.,][]{Morgan2003,Lesniewska2019}. At these redshifts ($z \gtrsim 6$), metals and dust grains must therefore mostly come from other stellar sources like SNe and SNRs, or interstellar grain growth processes. Note however that both theoretical and observational studies have nuanced this idea, suggesting that AGB stars may play a role at earlier times than usually assumed \citep[e.g.,][]{Valiante2009, Boyer2017}.

The impact of SNRs on the ISM dust budget of their host galaxy is dependent on the composition and size of the dust grains they form. Although observational evidence for the formation of dust grains in the ejecta of SNRs is increasing, it is also clear that the powerful blast from the initial explosion and its reverse shock destroys a significant proportion of the grains along the way. Theoretical studies have found that the surviving fraction of dust depends on the nature of grains, e.g. carbon-rich vs silicate-rich grains, as well as the size of grains \citep{Bianchi&Schneider2007, Nozawa2007, Biscaro2016, Micelotta2016}, and other environmental properties like the explosion energy, ambient ISM density and shock velocities and densities. Large grains ($\gtrsim 0.1~\mu$m) are thought to be more resilient to dust destruction by sputtering \citep{Silvia2010, Slavin2020}. However, studies that consider the combined effect of (thermal and kinetic) sputtering and grain-grain collisions find that large grains can be efficiently destroyed if relatively high density contrasts between clump and inter-clump media are considered \citep{Kirchschlager2019}. Carbonaceous grains appear able to survive reverse shock processing more efficiently than silicate-type of grains \citep{Silvia2010, Bocchio2016, Kirchschlager2019}, with iron grains showing even higher dust survival rates \citep{Silvia2010}. To understand the mass of freshly condensed supernova dust that is eventually mixed with the surrounding ISM material, it is thus crucial to understand the composition and average size of dust grains in SNRs.

Several recent studies have argued for the presence of relatively large dust grain radii ($> 0.1~\mu$m up to several microns) in several SNRs through the modelling of the observed dust continuum emission across infrared (IR) and sub-millimetre wavebands \citep{Gall2014, Fox2015, OwenBarlow2015, Wesson2015, Priestley2020} and -- in an independent way -- from modelling the red-blue asymmetries observed in late-time SN optical line profiles \citep{Bevan&Barlow2016, Bevan2017, NiculescuDuvaz2021}. 

The composition of supernova dust grains, however, remains relatively unconstrained. The presence of emission features of 100--500~K dust detected in mid-IR spectra suggests the presence of silicate-type grains (Mg$_{0.7}$SiO$_{2.7}$ or SiO$_{2}$) in (at least) two Galactic SNRs (Cas~A and G54.1+0.3; \citealt{Rho2008, Temim2017, Rho2018}) and a couple of extragalactic SNe (SN2004et and SN2005af; \citealt{Kotak2009, Fabbri2011, Szalai2013}). However, the grain composition of the bulk cold dust mass has been shown to differ from these mid-IR identified grain species in Cas~A based on post-explosion elemental abundance arguments \citep{DeLooze2017}, demonstrating that the warm dust observed with \textit{Spitzer} is not necessarily representative of the dominant supernova dust composition. The situation is more unclear for the majority of SNRs without spectroscopic data. The \textit{James Webb Space Telescope} will soon allow us to probe the onset of dust formation in young SNe and to characterise the composition of the dust formed during the first years post-explosion. 
But the absence of distinct dust emission or absorption features in the far-IR and sub-millimetre wavelength regimes will continue to hamper our ability to characterise the supernova dust composition. 
Further observations, in particular polarisation observations, can help break degeneracies and determine dust grain properties \citep[][]{Hensley2019WP}.
We therefore exploit multi-waveband polarimetric observations in this paper to constrain the bulk composition of supernova dust in the pulsar wind nebula (PWN), the Crab Nebula.

The Crab Nebula is believed to originate from a supernova type II-P \citep[][]{MacAlpineSatterfield2008}, although recent evidence suggests that it may have originated from an electron capture SN (ECSN) similar to the one observed for SN 2018zd \citep[][]{Hiramatsu2021}.  At a distance of 2~kpc \citep[][]{Trimble1968}\footnote{A recent estimate from \citet[][]{Fraser2019} with \textit{Gaia} places the Crab at 3.37~kpc, in which case the masses derived in this paper would be increased by a factor or $~\sim 2.8$.}, its proximity reveals stunning details at all wavelengths. Its synchrotron emission is observable from X-ray down to the radio regime \citep[e.g.][]{Hester2008}. Optical and UV spectra reveal a collection of emission lines, and its infrared emission exhibits clear filamentary dust structures. 
Despite extensive studies and its coverage from a large number of observing facilities, consensus has been hard to reach on some properties of the Crab, such as its dust mass. 

Early observations with IRAS, ISO and \textit{Spitzer} estimated dust masses ranging from 0.001 to 0.07~\msol \citep[][]{Marsden1984, Green2004, Temim2006, Temim2012}. Longer wavelength observations from \textit{Herschel}, combined with near-IR to radio observations to correct for synchrotron contamination, retrieved global dust masses of $0.11\pm0.01$~\msol  to $0.24^{+0.32}_{-0.08}$~\msol of carbonaceous or silicate dust, respectively \citep[][]{Gomez2012}. Using a physical model for the radiative heating of dust in the Crab, \citet[][]{TemimDwek2013} inferred a higher average dust temperature than previous work ($T_{\text{dust}}=56\pm2$~K), leading to a dust mass of $0.019^{+0.010}_{-0.003}$~\msol for a different type of carbonaceous grains. Combining photo-ionisation and dust radiative transfer models, \citet[][]{OwenBarlow2015} estimated 0.18--0.27~\msol of clumped amorphous carbonaceous dust, or a mix of 0.11--0.13~\msol and 0.39--0.47~\msol of amorphous carbonaceous and silicate dust, respectively. \citet[][]{DeLooze2019} were the first to correct for synchrotron emission and interstellar dust emission on a spatially resolved scale, which led to an inference of carbon-rich dust masses of 0.032--0.049~\msol of $T_{\text{dust}}=41\pm3$~K.
Replacing carbon- by common silicate-type grains (e.g., enstatite, MgSiO$_3$) slightly alters this dust mass estimate. Significantly higher dust mass estimates are obtained only for less emissive Fe or Mg$_{0.7}$SiO$_{2.7}$ grains — however, these high dust masses would violate the yields predicted by nucleosynthesis models \citep[e.g.][]{WoosleyWeaver1995}. In an independent way, \citet[][]{Nehme2019} inferred similar masses ($0.06\pm0.04$~\msol) of dust for the Crab Nebula. Using a physical model for radiative and collisional heating of the Crab’s dust, \citet[][]{Priestley2020} inferred a consistent carbonaceous dust mass of $\sim0.05$~\msol. Table~\ref{TabCrabMasses} summarises the dust mass estimates found in literature.

The uncertainties on the inferred dust masses for the Crab Nebula are partially driven by methodology, although the latest measurements seem to converge to somewhat lower dust masses, consistent with the estimated progenitor mass \citep[8--11~\msol;][]{MacAlpineSatterfield2008, Smith2013} and a modest condensation efficiency of around 10~per~cent. The unknown dust mixture in the Crab only adds to these uncertainties. Gas-phase abundance estimates of the Crab’s ejecta C/O ratios are above unity \citep[][]{Satterfield2012,OwenBarlow2015} suggesting that the Crab is carbon-rich. These carbon abundances contradict current CCSN nucleosynthesis models; but a revision in light of a possible ECSN scenario may be warranted. The ejecta also hosts several regions with C/O ratios below unity \citep[][]{DavidsonFesen1985, MacAlpineSatterfield2008} which suggests that carbon- and silicate-type grains are likely to co-exist within a single SNR. There appears to be less ambiguity on the Crab’s grain size distribution with the presence of micron-sized grains inferred through independent efforts to model the far-IR dust SED \citep[][]{TemimDwek2013, OwenBarlow2015, Priestley2020}. The observational inferences, however, contradict with theoretical model calculations of dust formation in PWN \citep[][]{Omand2019}, that suggest the presence of predominantly small ($<0.01~\mu$m) grains. 

In this paper, we constrain the carbon-to-silicate ratio in three dusty filaments of the Crab Nebula based on far-IR polarisation observations obtained with the HAWC+ instrument \citep[][]{Harper2018} on-board the Stratospheric Observatory for Infrared Astronomy airborne space observatory \citep[SOFIA][]{Temi2018}.
We derive silicate and carbonaceous grain temperatures, ranging from $\sim 30$ to $\sim 70$~K, and masses, ranging from $\sim 10^{-4}$ to $\sim 10^{-1}$~\msol, using 89 and 154~$\mu$m total and polarised intensities, and polarisation fractions.
Knowledge on the dust composition and properties in the Crab Nebula -- and SNRs in general -- is vital to estimate the net SN dust production rate and to assess the importance of SNRs in building up galactic dust budgets. 

Section~\ref{SecData} describes the data used in this study, while Section~\ref{SecResults} briefly focuses on the observed polarisation fractions in the Crab Nebula. After reviewing the removal of the synchrotron contribution in Section~\ref{SecSynchSub}, we derive dust properties (dust mass and temperature) in regions of the Crab in Section~\ref{SecConstraints}, and discuss the implications and limits of our results in Section~\ref{SecDiscussion}.
    
\section{Data}
\label{SecData}
The Crab Nebula was observed with the HAWC+ instrument \citep[][]{Harper2018}, on-board the Stratospheric Observatory for Infrared Astronomy \citep[SOFIA;][]{Temi2018}, in 2018 September (Proposal ID: 06\_0193; PI: Ilse De~Looze).
Polarimetric measurements were taken in bands C~(89~$\mu$m, FWHM~$\sim 7.8''$, pixel size $\sim 1.95''$) and D~(154~$\mu$m, FWHM~$\sim 13.6''$, pixel size $\sim 3.40''$).
The observations were done using the standard Nod Match Chop mode, with a total on-source exposure time of 1094.208~s (0.30~hrs), a chop frequency of 10.20~Hz and an amplitude of 200$''$. The dithering strategy was kept to default settings with four dithers and $20''$ scale. The data were processed with the {\tt HAWC\_DRP} pipeline, version 2.0.0 to Level~4, which includes the polarisation vectors, as described by the Data Product Handbook\footnote{\url{https://www.sofia.usra.edu/sites/default/files/Instruments/HAWC_PLUS/Documents/hawc_data_handbook.pdf}}.

\begin{figure*}
    \centering
    \includegraphics[width=\textwidth, clip, trim={2.5cm 1cm 3cm 0cm}]{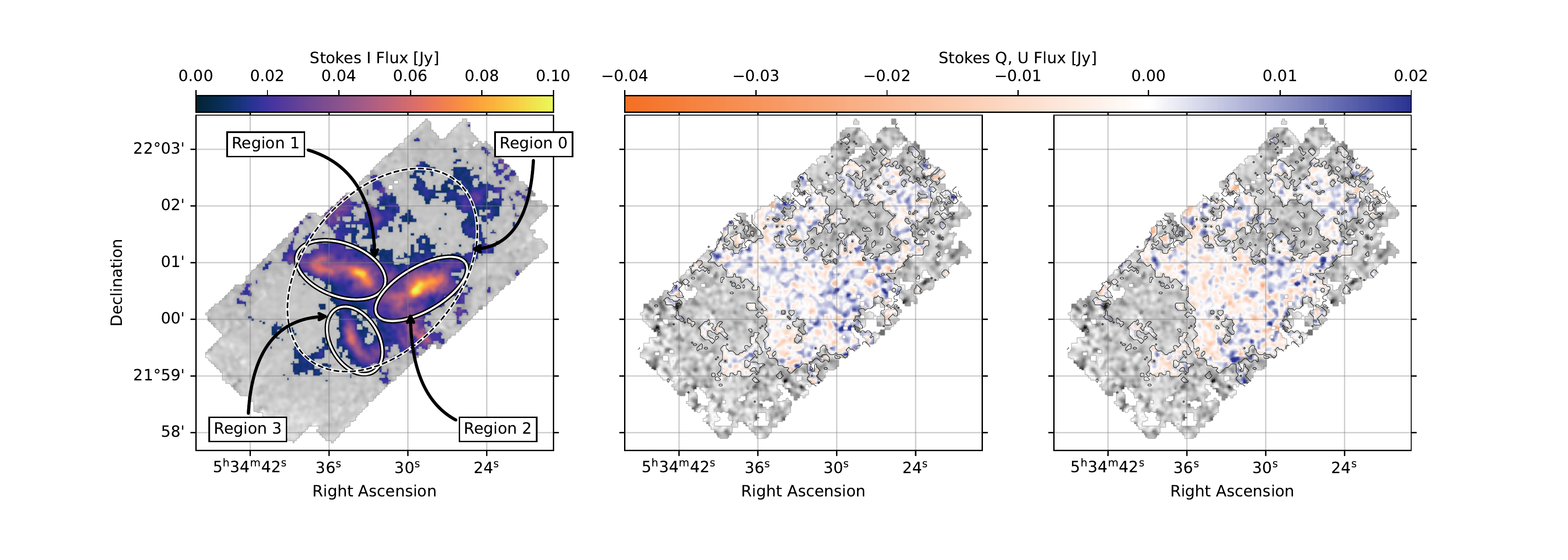}
    \includegraphics[width=\textwidth, clip, trim={2.5cm 1cm 3cm 0cm}]{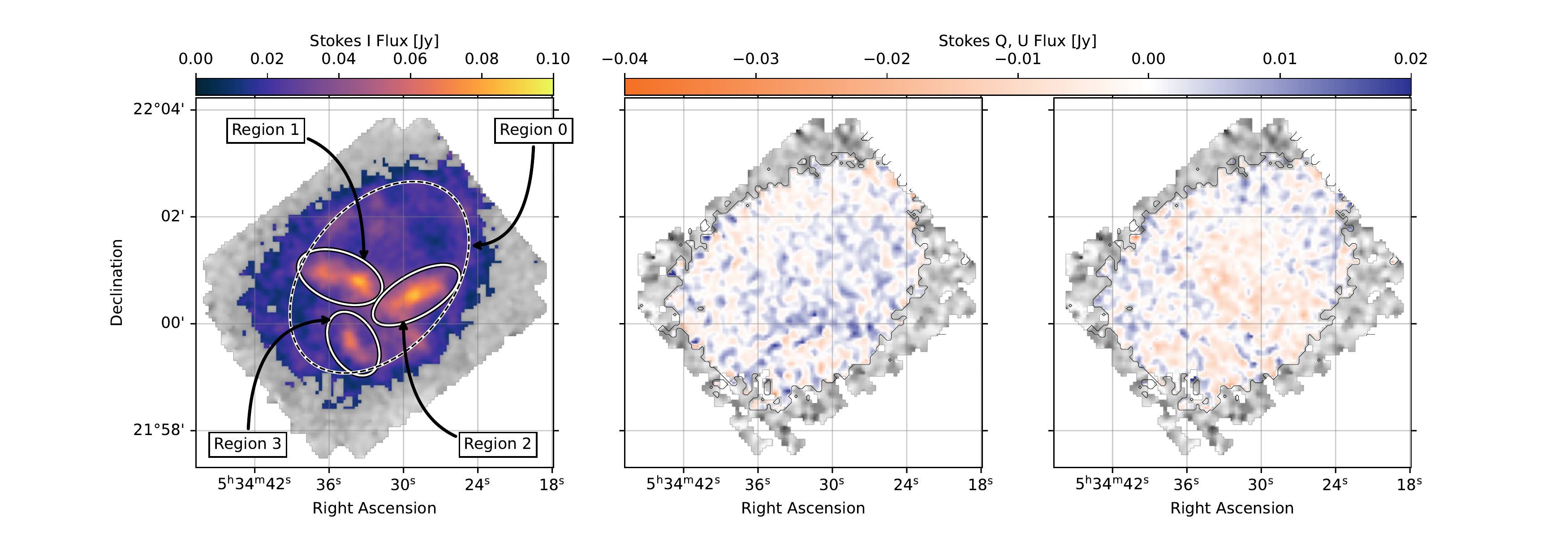}
    \caption{Native resolution SOFIA/HAWC+ data of the Stokes I (left), Q (middle) and U (right) parameters, in band C~89~$\mu$m ($\rm FWHM \sim 7.8''$, top) and band D~154~$\mu$m ($\rm FWHM \sim 13.6''$, bottom), in units of flux density. The background grey images show the entire field that was mapped. The coloured images show the data where $I/\sigma_{\rm I} \geq 3$.
    The white ellipses corresponds to three regions we focus on in the following sections, defined by eye, around dusty filaments. 
    The dashed-black ellipse is the region we use to compute integrated fluxes, and corresponds to half the one in \citet[][]{DeLooze2019}.}
    \label{FigDataStokes}
\end{figure*}

In Fig.~\ref{FigDataStokes}, we show the I, Q and U Stokes parameters in bands C~89~$\mu$m (top) and D~154~$\mu$m (bottom). For each band, the whole mapped area is shown in grey. The coloured maps show the pixels where $I/\sigma_I~\geq~3$.

\subsection{SOFIA Definitions}
\label{SecDefinitions}
The linearly polarised intensity, $I_{\rm pol}$, and its uncertainty, $\sigma_{I_{\rm pol}}$, are approximated as 
\begin{equation}
\begin{split}
    I_{\rm pol} &= \sqrt{Q^2 + U^2}, \\
    \sigma_{I_{\rm pol}} &= \sqrt{\frac{(Q~\sigma_Q)^2 + (U~\sigma_U)^2}{Q^2 + U^2}},
\end{split}
\end{equation}
with $\sigma_Q$ and $\sigma_U$ the uncertainties on Stokes vectors Q and U, respectively. The polarisation fraction (or polarisation degree) is the ratio of polarised intensity to Stokes I parameter. In the SOFIA framework, the \emph{debiased} polarisation, $p$, and its uncertainty, $\sigma_p$, are defined as
\begin{equation}
\begin{split}
    p &= \frac{1}{I}\times\sqrt{Q^2 + U^2 - \sigma^2_{I_{\rm pol}}}, \\
    \sigma_p &= p \times \sqrt{ \left ( \frac{\sigma_{I_{\rm pol}}}{I_{\rm pol}} \right )^2 + \left ( \frac{\sigma_I}{I} \right )^2 },
\end{split}
\label{EquPolFrac}
\end{equation}
(at high signal-to-noise) where $\sigma_I$ is the uncertainty of the Stokes I parameter. The polarisation angle, $\theta_{\rm p}$, and its uncertainty, $\sigma_{\theta_{\rm p}}$, are defined as 
\begin{equation}
    \begin{split}
        \theta_{\rm p} &= 0.5 \times {\rm arctan}(U/Q), \\
        \sigma_{\theta_{\rm p}} &= 0.5 \times \frac{ \sqrt{(Q~\sigma_U)^2 + (U~\sigma_Q)^2} }{Q^2 + U^2}
    \end{split}
\label{EquPolAng}
\end{equation}
with Q and U referenced to equatorial north.

\subsection{Modified Asymptotic Estimator}
\label{SecMAS}
Given the low signal-to-noise (S/N) of our HAWC+ data, we use the modified asymptotic (MAS) estimator \citep[][]{Plaszczynski2014} to recalculate the debiased polarisation, $p_{\rm MAS}$. We direct the reader to the reference paper for a detailed description of the MAS. Here are the main equations we use:
\begin{equation}
\begin{split}
    q &= Q/I,~u=U/I \\
    p &= \sqrt{q^2 + u^2} \qquad {\rm and} ~\theta_p = 0.5 \times {\rm arctan}(u/q), \\
    p_{\rm MAS} &= p - b^2\frac{1-e^{-p^2/b^2}}{2p}, \\
    {\rm with~} b^2 &= \sigma_u^2 \cos^2(\theta_p) + \sigma_q^2 \sin^2(\theta_p), \\
    \sigma^2_p &= \sigma_q^2~\cos^2(\theta_p) + \sigma_u^2~\sin^2(\theta_p).
\end{split}
\label{EquMAS}
\end{equation}

This estimate, also used in the data reduction pipeline of the James Clerk Maxwell Telescope SCUBA-2/POL2\footnote{\url{https://www.eaobservatory.org/jcmt/2020/04/new-de-biasing-method-for-pol2-data/}}, improves the calculation of the polarisation at low signal-to-noise ratios.  
The computation of the polarisation angle remains unchanged. 
In the following, when the polarisation fraction is given for a region of pixels, we first sum the intensities in Stokes I, Q, and U, and sum in quadrature the associated uncertainties. We apply the MAS estimator calculations to these summed intensities.
For simplification, from here on, we refer to the MAS estimate of the debiased polarisation simply as $p$.

No polarisation calculation is done using the SOFIA equations presented in Section~\ref{SecDefinitions}. However, the values listed in Table~\ref{TabMeanPol} are not affected whether we use the SOFIA or MAS approach: the latter provides polarisation at very low S/N, and using the former approach simply removes a few polarisation vectors in the maps. As we work with integrated values, the results are not affected by the chosen calculation.

\subsection{Ancillary data}
\label{SecAncData}
We use data products from \citet[][]{DeLooze2019}, particularly their best-fit parameters for the synchrotron radiation. We interpolate the synchrotron emission in bands C and D using the transmission curves\footnote{\url{https://www-sofia.atlassian.net/wiki/spaces/OHFC1/pages/1147682/7.+HAWC\#7.1.2-Performance}}. 

To compute the polarisation fraction and angle associated with that same synchrotron radiation, $p_{\rm radio}$ and $\theta_{\rm p_{radio}}$, we use the NIKA polarisation maps of the Crab Nebula at 150~GHz (FWHM~$\sim~18''$) presented in \citet{Ritacco2018}. 
The NIKA camera operated on the IRAM 30~m~telescope between 2012 and 2015, and observed in total intensity and polarisation. We refer the reader to the works of \citet[][]{Monfardini2010, Monfardini2011}, \citet[][]{Bourrion2012}, \citet[][]{Calvo2013}, and \citet[][]{Catalano2014} for a detailed description of the NIKA camera. Additionally, specifics about the polarisation calibration can be found in \citet[][]{Ritacco2017}.

\subsection{Image preparation}
Our synchrotron estimation in bands C and D is made at the SPIRE~500 resolution (FWHM~$\sim 36''$), from the \citet[][]{DeLooze2019} products. We convolve and regrid the HAWC+ data  to the SPIRE~500 resolution and pixel size ($\sim 14''$).

For both convolution and regridding, we use the \texttt{mpdaf} package\footnote{\url{https://mpdaf.readthedocs.io/en/latest/}} \citep[][]{Bacon2016}.
Convolution is done using simple 2D Gaussian kernels, with the target resolution being $36''$. 
Regridding is done using the \texttt{obj.Image.align\_with\_image} function, which allows for appropriate uncertainty propagation for variance maps, provided by the SOFIA pipeline.
We use the new maps of the Stokes parameters to calculate the debiased polarisation degree, the polarisation angle and their associated uncertainties using the modified asymptotic estimator in equation~(\ref{EquMAS}) on summed fluxes within each region, as explained in Section~\ref{SecMAS}.
In Fig.~\ref{FigPolRebin}, we show the convolved and rebinned Stokes I maps for HAWC+ C (left) and HAWC+ D (right) bands, with the polarisation vectors. The inset images show the S/N$_p$. Only a handful of pixels have values above 3. 
Note that the S/N ratios refer to the ratio of the intensity maps (for I, Q, and U) or percentage map (for $p$), with their associated error maps ($\sigma_X$). The errors in Stokes parameters are given by the SOFIA pipeline, while the error on $p$ is given in Section~\ref{SecMAS}. This leads to high values of S/N in total intensity, where the source is well detected, but larger errors in the Stokes Q and U parameters result in significantly lower S/N values on the $p$ values.

\subsection{Integrated fluxes}
We compute the total fluxes inside an elliptical aperture centred on the Crab pulsar (RA: 83\fdg633; Dec: 22\fdg0145), with a 50\degr rotation, and minor and major axes 122\farcs5x81\farcs5 (region 0, thick black-dashed ellipse in Fig.~\ref{FigDataStokes}). We find integrated fluxes for the SOFIA data of $136.9\pm3.4$ and $79.9\pm1.5$~Jy/pix in bands HAWC~C~89~$\mu$m and HAWC~D~154~$\mu$m, respectively.

The minor and major axes of the ellipse are half of those in \citet[][]{DeLooze2019}, because the coverage of the HAWC data is smaller than the one they used in their paper. For comparison, we recalculate the {\it Herschel} integrated fluxes from the data in \citet[][]{DeLooze2019}. The {\it Herschel} fluxes within this aperture are $128.8\pm10.3$, $120.1\pm9.6$, $85.3\pm6.8$, $48.8\pm4.9$, $49.3\pm4.9$, and $52.8\pm5.3$~Jy/pix, at PACS~70, PACS~100, PACS~160, SPIRE~250, SPIRE~350 and SPIRE~500, respectively\footnote{Assuming a constant Galactic ISM emission across the nebula, and 8~per~cent uncertainties for PACS data, and 10~per~cent for SPIRE data.} \citep[accordingly, about half the values reported by][]{DeLooze2019}. 
We report $p$ values within that region in Fig.~\ref{FigPlotpValues} and Table~\ref{TabMeanPol}.

\subsection{Morphology and polarisation}
\label{SecResults}
\begin{figure*}
    \centering
    \includegraphics[width=\textwidth, clip, trim={1.5cm 0 1.5cm 0.2cm}]{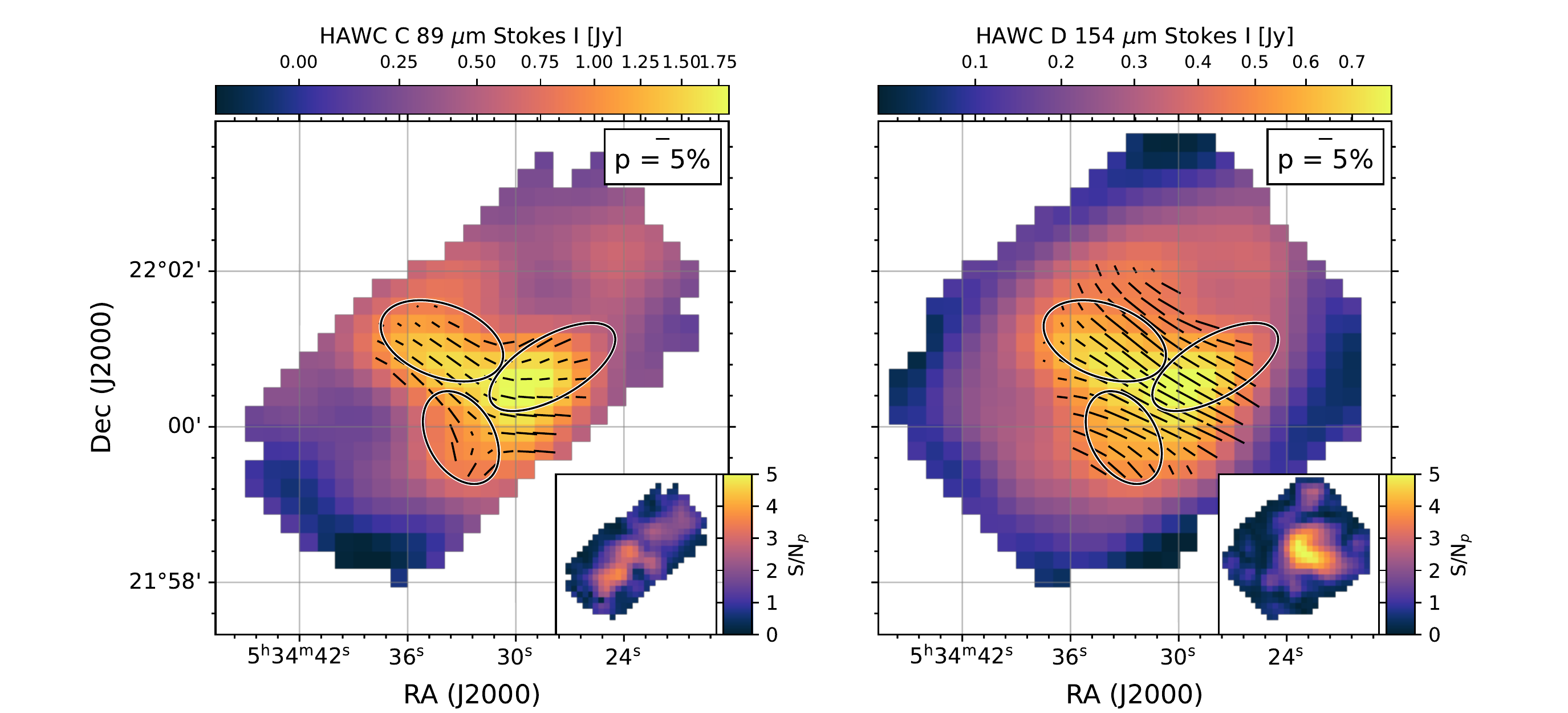}
    \caption{Stokes~I maps at the SPIRE~500 resolution ($36''$) and pixel size ($14''$) in band C (left) and D (right), with polarisation vectors where $I_{\rm Stokes~I} \gtrsim 0.9$ and $0.4$~Jy/pix, respectively. These thresholds are only used for display purposes and not for calculations. 
    The insets show the signal-to-noise in polarisation, S/N$_p$. A number of pixels in the regions of interest have low S/N$_p$, leading to integrating fluxes in the highlighted ellipses (Fig.~\ref{FigDataStokes}).}
    \label{FigPolRebin}
\end{figure*}

Fig.~\ref{FigDataStokes} shows the Stokes vectors at native resolution.
The dusty filaments appear clearly at native resolution, with the highest signal and S/N in Stokes~I, and are highlighted by the three regions we define: 
\{83\fdg646, 22\fdg014, 49\farcs216, 28\farcs115, 160\degree\}, 
\{83\fdg621, 22\fdg009, 54\farcs383, 23\farcs575, 30\degree\}, 
\{83\fdg642, 21\fdg994, 38\farcs607, 25\farcs531, 120\degree\}, for \{RA, Dec, major and minor axes, and rotation\}, for regions 1, 2 and 3, respectively.
Fig.~\ref{FigPolRebin} shows the polarisation vectors at SPIRE~500 resolution. Note that these polarisation vectors correspond to the observed polarisation angles rotated by 90\degree\ and represent the direction of the magnetic field.
We lose significant spatial information when convolving to a $36''$ point spread function. While filaments 1 and 2 are still clearly visible in intensity, the third region is largely smeared out and blended. The polarisation vectors do not clearly mark the filaments. 

The vectors shown in Figs.~\ref{FigPolRebin} and \ref{FigFinalPol} are only for display purposes, but do not have sufficient S/N in the Stokes parameters to be reliable for a resolved analysis. We choose thresholds of $I_{\rm Stokes~I} \sim 0.9$ and $0.4$~Jy/pix in bands C and D, respectively, only to display polarisation vectors (which correspond to ${\rm S/N_{I} \sim 40}$ in both bands).

\begin{figure}
    \centering
    \includegraphics[width=0.5\textwidth, clip, trim={1cm 0.75cm 1cm 0.75cm}]{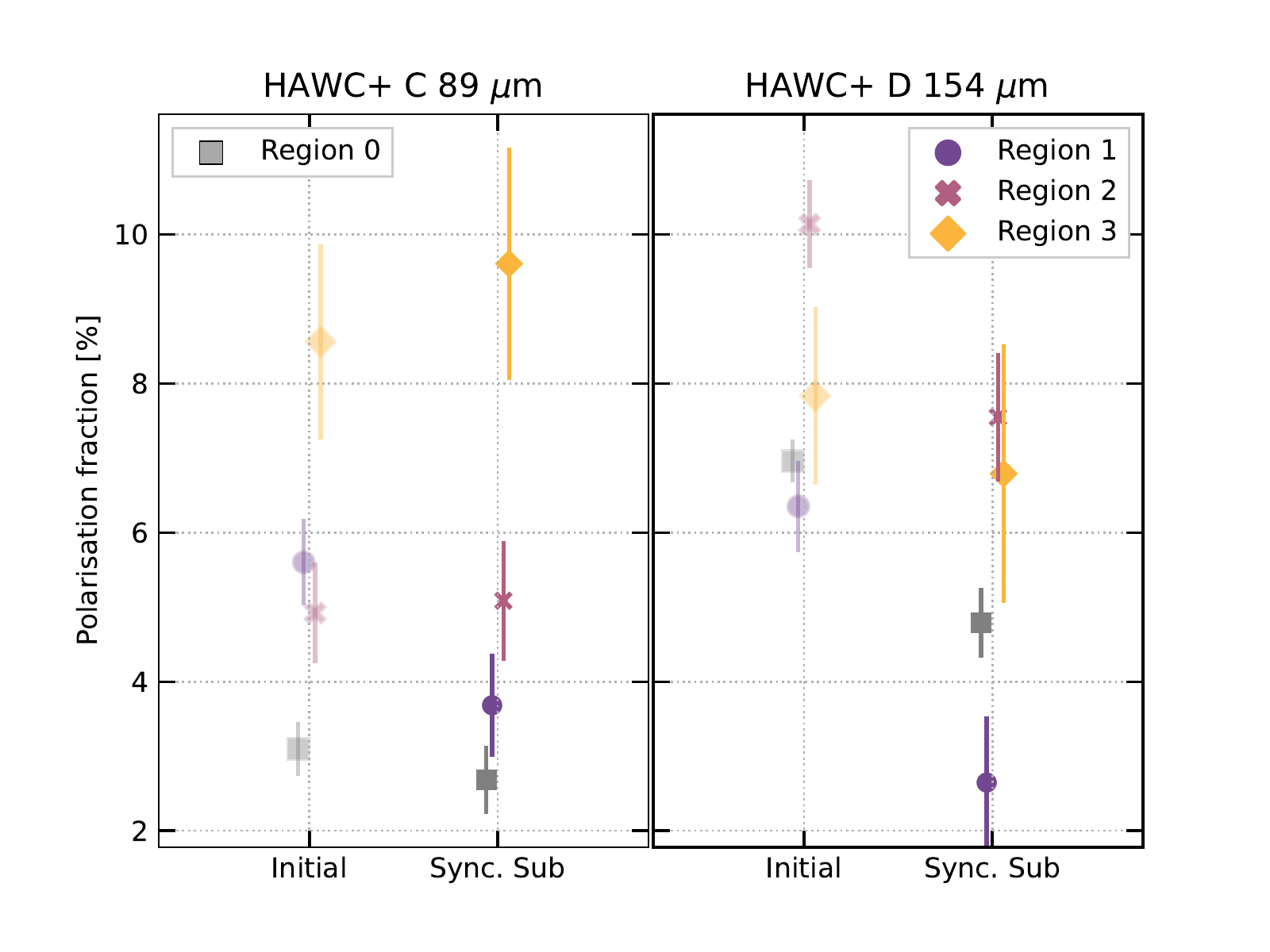}
    \caption{Variations of the polarisation fraction in the three regions highlighted in this paper and the larger region, in bands C and D, before and after synchrotron subtraction at the SPIRE~500 resolution (tabulated in Table~\ref{TabMeanPol}). The error bars represent the 1-$\sigma$ error on polarisation, $\sigma_p$, as described by equation~(\ref{EquMAS}).}
    \label{FigPlotpValues}
\end{figure}

In Fig.~\ref{FigPlotpValues}, we report the values of the polarisation fraction $p$ after convolution and projection to the SPIRE~500 resolution and pixel grid, in the three regions described above. The same values are tabulated in Table~\ref{TabMeanPol}.
The average polarisation in each region never exceeds 10~per~cent, and can be as low as $\sim 2.5$~per~cent. This is (much) lower than previously detected polarisation in Cas~A by \citet[][]{Dunne2009}. Note that the number of pixels within each region at the SPIRE~500 resolution is fairly low (between 16 and 23 pixels).

\renewcommand{\arraystretch}{1.2}
\begin{table}
    \centering
    \begin{tabular}{c|l c c}
         & & {\textbf{HAWC+ C}} & {\textbf{HAWC+ D}} \\
         \hline
                                & Reg. 0 & $3.1 \pm 0.4$ & $ 6.7\pm 0.3$ \\ 
            Initial data        & Reg. 1 & $ 5.6\pm0.6 $ & $ 6.4\pm0.6 $ \\
                                & Reg. 2 & $ 4.9\pm0.7 $ & $ 10.1\pm0.6 $ \\
                                & Reg. 3 & $ 8.6\pm1.3 $ & $ 7.8\pm1.2 $\\
        \hline
                                & Reg. 0 & $2.7 \pm 0.5$ & $ 4.8\pm 0.5$ \\ 
            Synchrotron         & Reg. 1 & $ 3.7\pm0.7 $ & $ 2.7\pm0.9 $ \\
            subtracted          & Reg. 2 & $ 5.1\pm0.8 $ & $ 7.6\pm0.9 $ \\
                                & Reg. 3 & $ 9.6\pm1.6 $ & $ 6.8\pm1.7 $ \\   
        \hline
    \end{tabular}
    \caption{Polarisation degrees and associated errors, $p \pm \sigma_p$, in per cent, as calculated in equation~(\ref{EquMAS}). The values are given after integrating fluxes (and errors) within each region, at the SPIRE~500 resolution.}
    \label{TabMeanPol}
\end{table}

\section{Synchrotron radiation removal}
\label{SecSynchSub}
The resolved synchrotron emission in the Crab was fit by \citet[][]{DeLooze2019} at the SPIRE~500 resolution, using a broken power-law. The normalisation factor, $F_{\nu_0}$, IR power-law index, $\alpha_{\rm IR}$, and wavelength of the break, $\lambda_{\rm b}$, are free parameters, and the radio power-law index is fixed (0.297):
\begin{equation}
    F_\nu = 
    \begin{dcases}
    F_{\nu_0} \times \left ( \frac{\nu}{\nu_0} \right )^{-0.297} \qquad &{\rm if \lambda \geq \lambda_{b}}\\
    F_{\nu_0} \times \left ( \frac{\nu}{\nu_0} \right )^{-\alpha_{\rm IR}} \times \left ( \frac{\nu_{\rm b}}{\nu_0} \right )^{-0.297} \times \left ( \frac{\nu_{\rm b}}{\nu_0} \right )^{\alpha_{\rm IR}} \qquad &{\rm otherwise}
    \end{dcases}
\end{equation}
These authors found that the contribution of the synchrotron radiation can be significant and contribute up to 19, 23 and 35 per cent, in the 70, 100, and 160~$\mu$m PACS bands, respectively. 
\begin{figure*}
    \centering
    \includegraphics[width=\textwidth, clip, trim={0.5cm 0 0.5cm 0}]{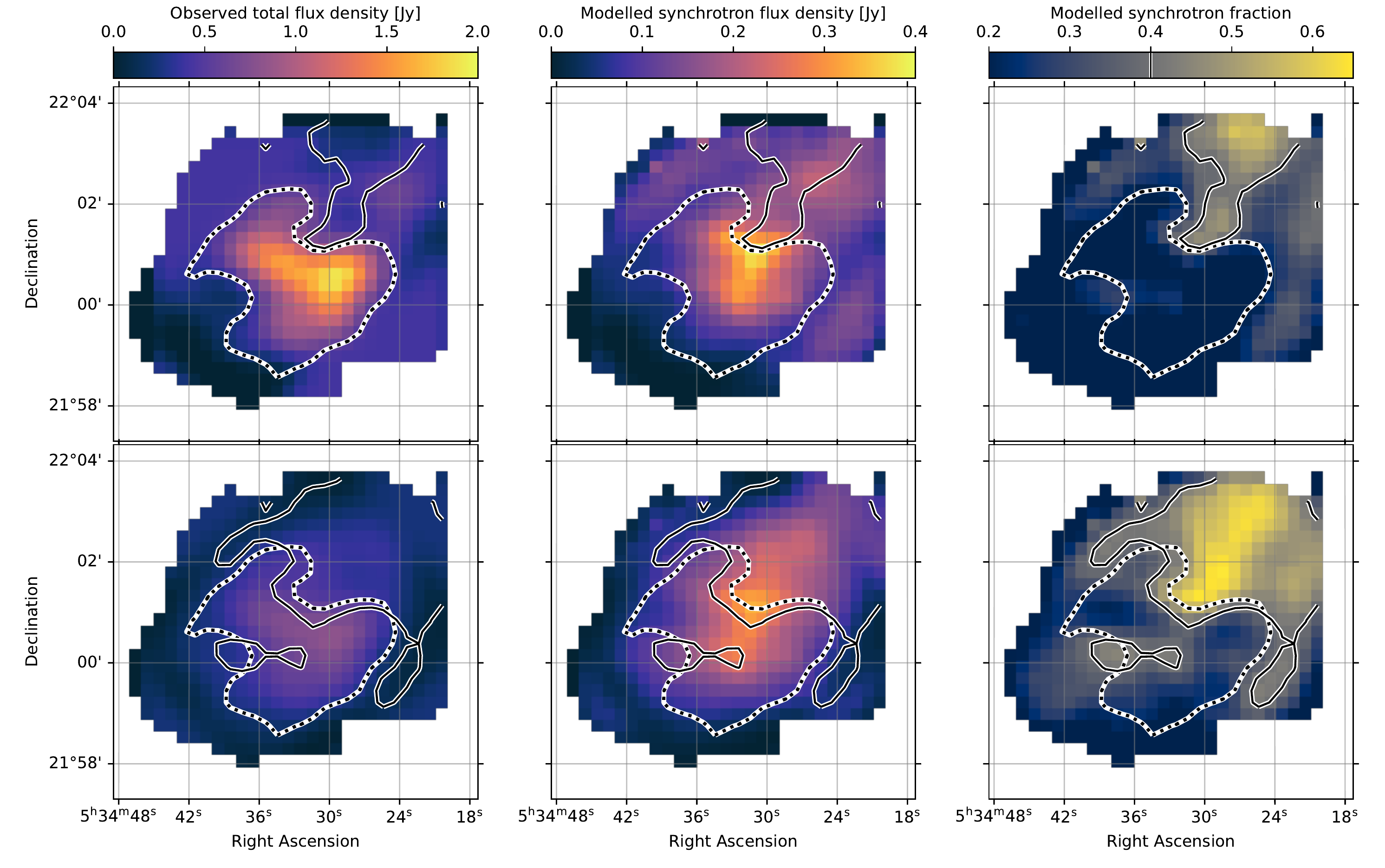}
    \caption{\textit{Top:} HAWC+ C band. \textit{Bottom:} HAWC+ D band. \textit{From left to right:} Total intensity convolved and regridded to the SPIRE~500 resolution and pixel grid, estimated synchrotron intensity, and fraction of the synchrotron emission with respect to the total intensity. These maps were derived from the synchrotron best fits in \citet[][]{DeLooze2019} and interpolated to the HAWC+ bands. We use these maps to compute the synchrotron total intensity at 89 and 154~$\mu$m, $I_{\rm sync}$. The solid contours mark the limits where the synchrotron fraction reaches 40~per~cent. The dotted contours mark the limits of ${\rm M_{dust} = 0.0001}$~\msol per~pixel in \citet[][]{DeLooze2019}.}
    \label{FigfSync}
\end{figure*}
Using their best fit models that include the synchrotron radiation, we create predicted synchrotron emission maps at 89 and 154~$\mu$m, and the corresponding synchrotron fraction of total flux, $f_{\rm sync}$, at the SPIRE~500 resolution. Based on the \citet[][]{DeLooze2019} fits, the error associated with $f_{\rm sync}$ is about 5~per~cent in the regions of interest.
Fig.~\ref{FigfSync} shows from left to right, the HAWC+ data, the synchrotron intensity, and the synchrotron fraction, in the C (top) and D (bottom) bands. 

\begin{figure*}
    \centering
    \includegraphics[width=\textwidth, clip, trim={1.5cm 0 1.25cm 0.2cm}]{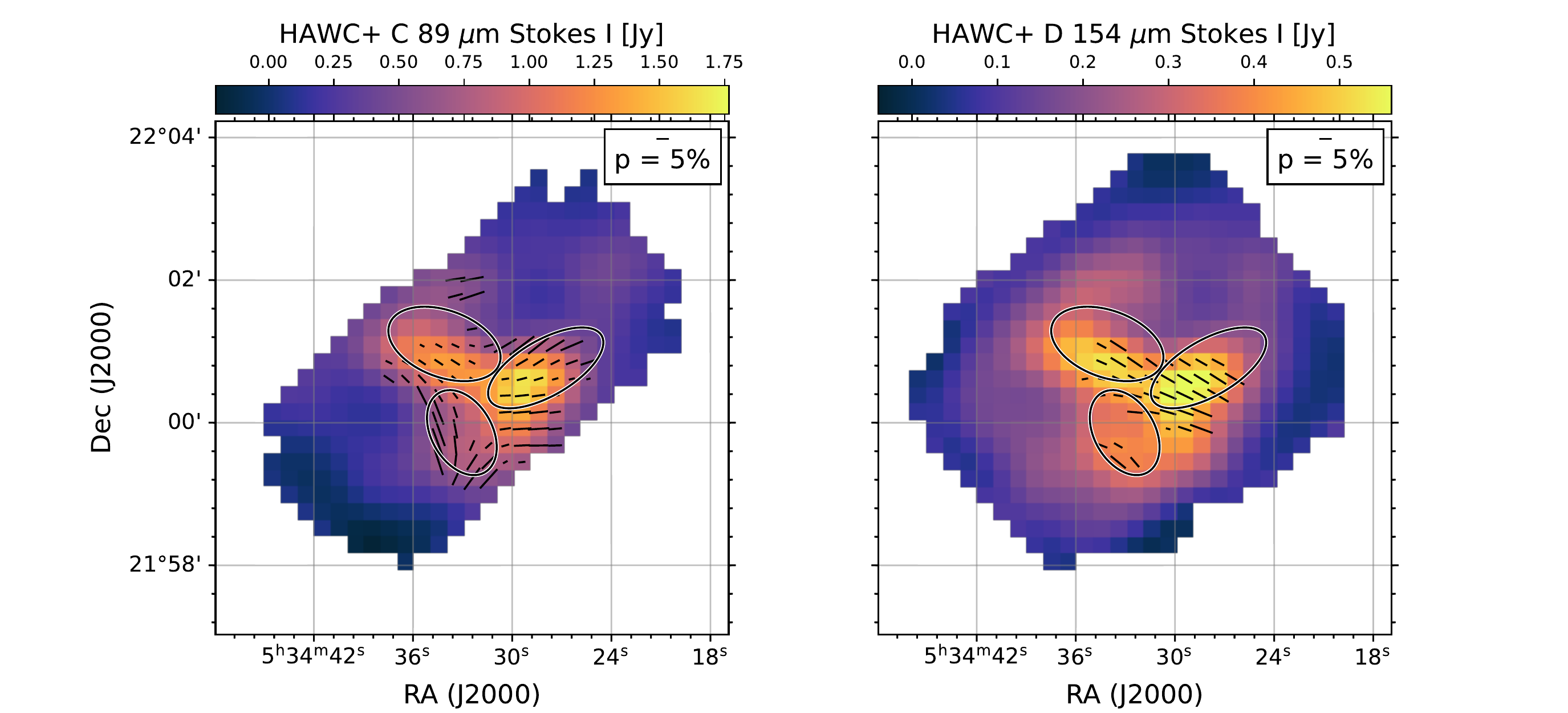}
    \caption{Synchrotron-subtracted Stokes~I maps in the HAWC+ C (left) and HAWC+ D (right) maps, with the synchrotron-free polarisation vectors overlaid (with polarisation vectors where $I_{\rm Stokes~I} \gtrsim 0.5$ and $0.4$~Jy, respectively, thresholds only used for display purposes and not calculations).
    See Appendix~\ref{AppOverplots} for a comparison between the synchrotron subtracted images (the top row images are the same as this Figure), the NIKA~150~GHz image, and the total dust mass from \citet[][]{DeLooze2019}.}
    \label{FigFinalPol}
\end{figure*}

We correct the observed HAWC+ Stokes I, Q and U maps after estimating the synchrotron emission, $I_{\rm sync} = f_{\rm sync} \times I_{\rm HAWC}$, and calculating the associated $Q_{\rm sync}$ and $U_{\rm sync}$ Stokes parameters, following:
\begin{equation}
    \begin{split}
        P_{\rm sync} &= p_{\rm radio} \times I_{\rm sync} \\
        Q_{\rm sync} &= P_{\rm sync} \times {\rm cos}(2~\theta_{\rm p_{radio}}) \\
        U_{\rm sync} &= P_{\rm sync} \times {\rm sin}(2~\theta_{\rm p_{radio}}).
    \end{split}
\end{equation}
where $p_{\rm radio}$ and $\theta_{\rm p_{radio}}$ are the polarisation fraction and angle at radio wavelengths (dominated by synchrotron), from the NIKA polarisation maps presented in Section~\ref{SecAncData} (calculated using the MAS estimator presented in equation~(\ref{EquMAS}) and the polarisation angle using equation~(\ref{EquPolFrac})). We assume that the synchrotron polarisation fraction is constant across frequencies \citep[][]{Ritacco2018}.
We make sure to use $\theta_{\rm p_{radio}}$ in equatorial coordinates for this purpose.
We can then estimate the final linear polarisation without the synchrotron contribution simply as follows:
\begin{equation}
    \begin{split}
        I_{\rm dust} &= I_{\rm HAWC+} - I_{\rm sync}  \\
        Q_{\rm dust} &= Q_{\rm HAWC+} - Q_{\rm sync}  \\
        U_{\rm dust} &= U_{\rm HAWC+} - U_{\rm sync}. \\
    \end{split}
\end{equation}
We propagate errors accordingly, using initial uncertainties on $f_{\rm sync}$ from the modelling in \citet[][]{DeLooze2019}, and $p_{\rm radio}$ and $\theta_{\rm radio}$ from the observations in \citet[][]{Ritacco2018}. Note that most of the errors on the final polarisation come from the quality of the HAWC+ Stokes vector.
Fig.~\ref{FigFinalPol} shows the maps of the Stokes~I parameter after subtracting the synchrotron contribution in the HAWC+ C and D bands, with the polarisation vectors. For comparison, Fig.~\ref{AppPanels} shows the same maps with the NIKA~150~GHz Stokes I parameter and the dust mass map from \citet[][]{DeLooze2019}.
The synchrotron-free polarisation fractions are reported in Table~\ref{TabMeanPol} and plotted in Fig.~\ref{FigPlotpValues}.
In band~D, the polarisation decreases in all regions after removal of the synchrotron radiation. In band~C, the average value increases in regions 2 and 3.
The fairly low S/N makes it difficult to truly interpret the variations before and after synchrotron subtraction, and between bands.
Similarly to Fig.~\ref{FigPolRebin}, the shown polarisation vectors are only for display purposes, but not used in the calculations. The thresholds used are pixels with $I_{\rm Stokes~I} \sim 0.5$ and $0.4$~Jy/pix in bands C and D, respectively (which corresponds to S/N$_{\rm I} \sim 9$ in both bands) in the synchrotron-subtracted maps.

\begin{figure*}
    \centering
    \includegraphics[width=\textwidth, clip, trim={1.4cm 1cm 2.5cm 0.75cm}]{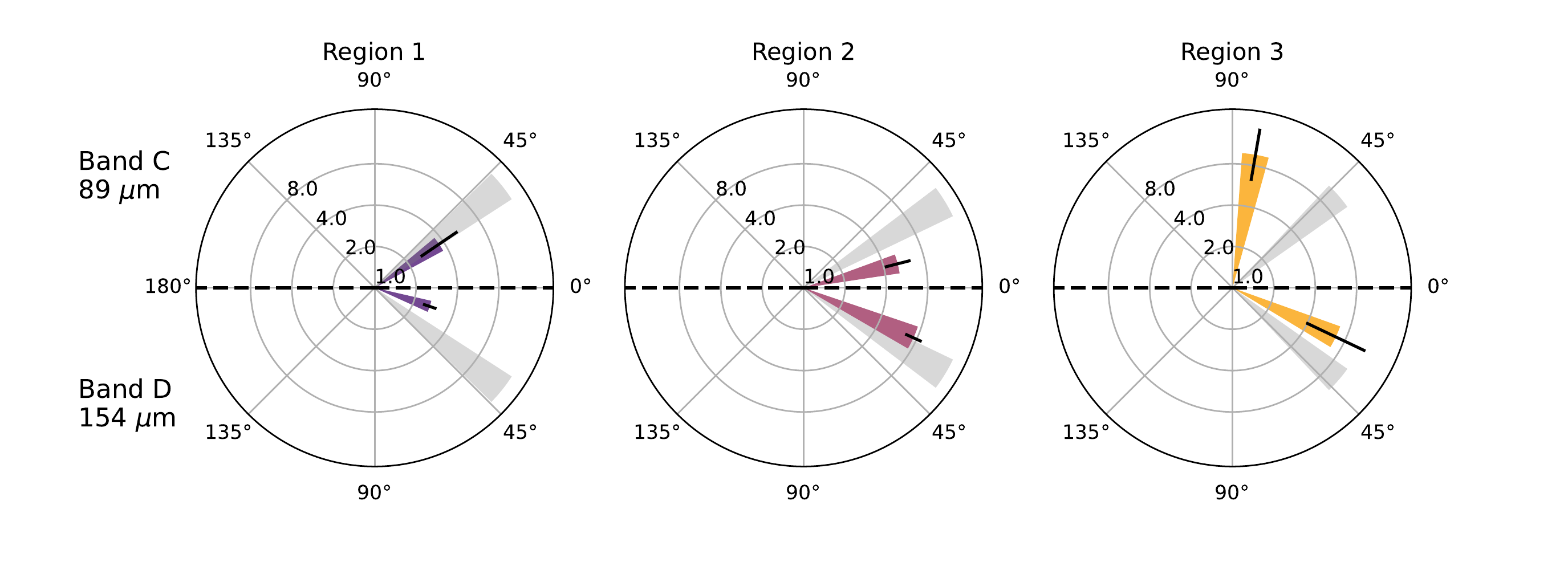}
    \caption{Polar representation of the polarisation vectors in each region (columns), in band~C (top half) and band~D (bottom half). The angle is the polarisation angle, and the radius is the polarisation fraction. The colour bars are the value of the dust-only polarisation vectors, while the grey bars are those from the NIKA data, at 150~GHz.}
    \label{FigFinalPolarPlot}
\end{figure*}

In Fig.~\ref{FigFinalPolarPlot}, we show the dust-only polarisation vectors (colour bars) in each region for bands~C (top half) and D (bottom half), with the NIKA~150~GHz polarisation vectors (grey bars) in the same regions. The plotted angle is the polarisation angle and the polar radius is the polarisation fraction. We can see that the angles for dust and synchrotron polarisation do not differ much, as expected if both are polarised perpendicular to local magnetic field lines.

\section{Constraints on dust}
\label{SecConstraints}
In this section, we derive dust properties in the Crab Nebula using the dust-only polarisation fractions, and the dust-only total intensity and polarised intensity maps. 

\subsection{Modelling assumptions}
\textit{On the dust --}
We assume that dust in the Crab Nebula is a mixture of dust grains that are either silicate-rich, hereafter bearing the subscript `Sil', or carbon-rich, hereafter bearing the subscript `aC'.
We parameterise their relative abundances using the fraction of carbonaceous grains, \fac, so that we can write
\begin{equation}
    \Sigma_{\rm aC} = f_{\rm aC}~\Sigma_{\rm dust} \qquad{\rm and}\qquad \Sigma_{\rm Sil} = (1-f_{\rm aC})~\Sigma_{\rm dust},
\end{equation}
where $\Sigma_X$ the mass surface density of $X=\{{\rm Sil, aC, dust}\}$.
We assume optically thin thermal emission from each component at a single steady-state temperature \citep[][]{LiDraine2001}, using its opacity law and a blackbody (as we use large grains, see later in the text):
\begin{equation}
    I_{X}(\lambda) = \kappa_{X}(\lambda) ~ \Sigma_X ~ B(\lambda, T_X)
\end{equation}
where $X = \{{\rm Sil, aC}\}$, $\kappa$ is the mass absorption coefficient, $T$ is the dust grain temperature, and $B(\lambda, T)$ is the Planck function.

\textit{On the carbonaceous grains --}
In this section, we assume that carbonaceous grains do not align with the magnetic field, and therefore do not contribute to the observed polarisation.
This assumption stems from the fact that no observation of the 3.4~$\mu$m C-H stretch absorption feature has conclusively detected polarised signal \citep[e.g.,][]{Adamson1999, Ishii2002, Chiar2006, Mason2007}.
In the following, we use a population of carbon-rich material with optical properties from the optECs data \citep[][]{Jones2012a, Jones2012b, Jones2012c} with E$_{\rm g}=0.1$~eV and $\rho_{\rm aC} = 1.6$~g~cm$^{-3}$.

\textit{On the silicate grains --}
Although there is no strict consensus on the minimum radius for a grain to polarise light, we assume that grains with radius $a \leq 0.1~\mu$m do not polarise. Effectively, smaller grains are able to polarise light but to a lower degree because of a low alignment efficiently \citep[e.g.][]{DraineFraisse2009, Guillet2018, DraineHensley2021StarPol}, and we do not consider them, for simplification purposes.
Additionally, we assume that all silicate grains align. This assumption simplifies greatly the equations, and facilitate their solving given the limited amount of data available.
We investigate different silicate materials, using their (polarised) absorption cross-sections. 
For this purpose, we use \costuum\footnote{\url{https://github.com/SKIRT/CosTuuM}} \citep[][]{Vandenbroucke2020}, an open-source C++-based Python library that computes infrared absorption cross-sections, $Q_{\rm abs}$, for a variety of grain compositions and grain sizes. 
For silicate material, we use the refractive indices, (n$_{\rm Sil}$, k$_{\rm Sil}$), and mass densities, $\rho_{\rm Sil}$, for five different types of silicates, downloaded from the JENA database\footnote{\url{https://www.astro.uni-jena.de/Laboratory/OCDB/index.html}}. 

\textit{On the grain size and shape distributions --}
We assume that both types of grains have the same size-distribution. Since there is no observable evidence of the existence of a reverse shock in the Crab Nebula \citep[e.g.,][]{Hester2008}, we do not consider that a particular grain species has undergone more processing than the other, and has a significantly different size distribution. We use a \citet[][]{MRN1977} power-law size distribution (index of 3.5), for grains with radius $0.1 \leq a \leq 5~\mu$m.
We fix the shape distribution to the ``continuous distribution of ellipsoids'' CDE2 model from \citet[][]{Ossenkopf1992, DraineHensley2021}. Note that the most extreme shapes in this distribution are not sampled in this work.

\textit{On the alignment mechanism and zenith angle --}
Assuming that alignment is due to the local magnetic field and the polarisation is parallel to the main axis of the grain (in the Rayleigh regime, $a\ll\lambda$), the observed polarisation is perpendicular to that magnetic field direction, in a plane perpendicular to the line-of-sight.
If we further assume that all the magnetic field vectors are in the same plane, we can use a fixed zenith angle (angle between magnetic field orientation and observer) that only depends on the inclination of the Crab Nebula.
Recent 3D modelling of the Crab seems to show that it has a relatively cylindrical/spherical structure \citep[][]{Martin2021}. Given the complexity of the 3D structure, we present the dust masses for a range of zenith angles $\theta$ (Fig.~\ref{FigFinalMasses}). We choose a range of 45 to 90\degree, representing a mix of random grain orientations and high alignment efficiency for all grains.
In each case, we assume all grains to have the same alignment angle for simplification, and the following equations therefore do not depend on $\theta$.

\subsection{Method}
\label{SecMethod}
We make use of the integrated values in the three regions (Section~\ref{SecResults}) of the observed dust-only polarised intensity, $I_{\rm pol}(\lambda)$, total intensity, $I(\lambda)$, and dust-only polarisation fraction $p(\lambda)$, in bands C and D, to derive silicate and carbonaceous grain temperatures, their masses, and the fraction of carbonaceous grains.

\textit{1. Silicate grain temperature, \tsil~--~}
Given the previous assumptions, we can write
\begin{equation}
    I_{\rm pol}(\lambda) = I_{\rm Sil, pol}(\lambda) = \kappa_{\rm Sil, pol}(\lambda) ~ \Sigma_{\rm Sil} ~ B(\lambda, T_{\rm Sil}).
\label{EquSolveTSil}
\end{equation}
We solve for \tsil using the ratio $I_{\rm Sil, pol}(89~\mu {\rm m})/I_{\rm Sil, pol}(154~\mu {\rm m})$, in each region.

\textit{2. Carbonaceous grain temperature, \tac~--~}
Given the previous assumptions, we can express the polarisation fraction, $p(\lambda)$, as 
\begin{equation}
    p(\lambda) = \frac{I_{\rm pol}(\lambda)}{I_{\rm tot}(\lambda)} = \frac{I_{\rm Sil, pol}(\lambda)}{I_{\rm aC}(\lambda) + I_{\rm Sil}(\lambda)}, \qquad {\rm  and,}
\label{EquSolveTaC}
\end{equation}
\begin{equation}
    I_{\rm tot}(\lambda) = \kappa_{\rm aC}(\lambda)~\Sigma_{\rm aC}~B(\lambda, T_{\rm aC}) + \kappa_{\rm Sil}~\Sigma_{\rm Sil}~B(\lambda, T_{\rm Sil}).
\end{equation}
We solve for \tac using the ratios between the C and D bands, and using the previously derived \tsil, in each region.

\textit{3. Fraction of carbonaceous grains~--~}
Using the observations and the dust temperatures inferred, we compute each population mass surface density, $\Sigma_{\rm Sil}$ and $\Sigma_{\rm aC}$, and solve for \fac:
\begin{equation}
    f_{\rm aC} = \Sigma_{\rm aC} / (\Sigma_{\rm aC} + \Sigma_{\rm Sil}).
\end{equation}
Note that solving for the dust surface densities is done by choosing either the fluxes in band C or band D, individually. 
The final results are given as the average of the $\Sigma_{\rm aC}$ and $\Sigma_{\rm Sil}$ derived using band C and band D separately.

\subsection{Results}
\label{SecPolResults}
\begin{figure}
    \centering
    \includegraphics[width=0.5\textwidth, clip, trim={0 0.75cm 1.5cm 0}]{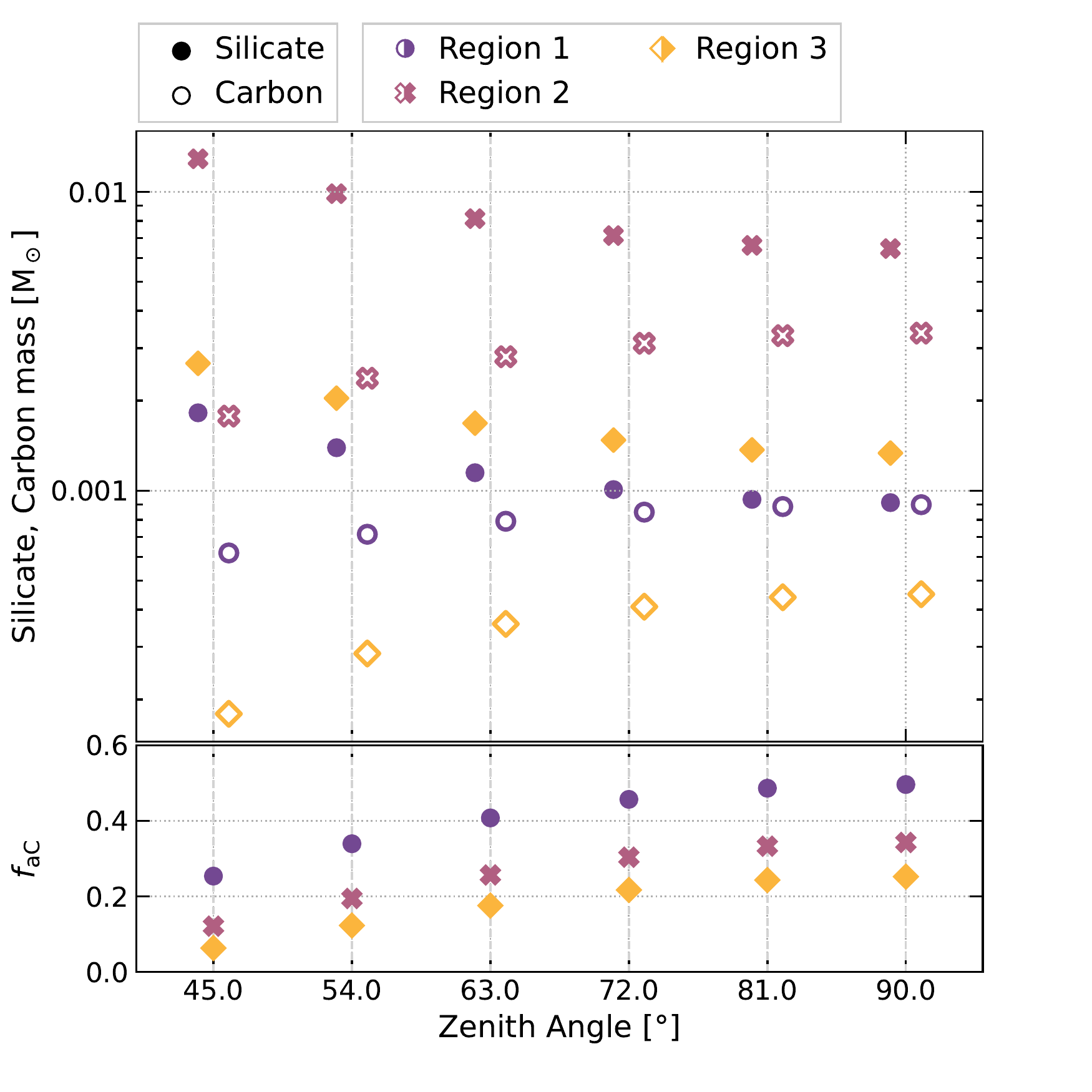}
    \caption{Final silicate (MgSiO$_3$, filled symbols) and carbonaceous (empty symbols) grain masses as a function of the zenith angle for $45\degree \leq \theta \leq 90\degree$. As expected, the silicate mass decreases with increasing $\theta$, leading to \fac increasing.}
    \label{FigFinalMasses}
\end{figure}

We report the carbonaceous and silicate grain temperatures and masses in Table~\ref{TabCrabMasses}, for a fixed carbonaceous grain composition, and five silicate materials: MgSiO$_3$, amorphous and glassy, Mg$_{0.5}$Fe$_{0.5}$SiO$_3$, and Mg$_{0.7}$SiO$_{2.7}$.
We choose MgSiO$_3$ (enstatite) and Mg$_{2}$SiO$_{4}$ (forsterite) because they are commonly used in other SNR-related studies \citep[e.g.][]{Nozawa2003, Sarangi2015, Sluder2018}, Mg$_{0.7}$SiO$_{2.7}$ because it provided good fits to the solid-state features observed in \textit{Spitzer}/IRS spectra of other SNRs \citep[e.g., Cas~A][]{Arendt2014}, and Mg$_{0.5}$Fe$_{0.5}$SiO$_3$ to test the inclusion of Fe on the derived dust properties.
Note that the dust masses M$_{\rm aC}$ and M$_{\rm Sil}$ move in opposite direction (Fig.~\ref{FigFinalMasses}): when considering the low (high) mass bound of M$_{\rm aC}$, it is paired with the high (low) mass bound of M$_{\rm Sil}$.
We calculate uncertainties on the temperatures by solving equations~(\ref{EquSolveTSil}) and (\ref{EquSolveTaC}) for a sample of values within range of fluxes (in polarised and total intensities) considering Gaussian uncertainties.
Note that in our framework, the derived dust temperatures are independent of the chosen zenith angle. We found that using angles from 45\degree\ to 90\degree, the ratios of ${\rm \kappa(89~\mu m)/\kappa(154~\mu m)}$, on which \tsil and \tac depend, only varies by $\sim 0.1$~per~cent. 

The dust temperatures roughly agree with previous works. Carbon-rich grains, with temperatures ranging from $\sim 40$~K to $\sim 70$~K, show higher temperatures than silicate grains, with temperatures ranging from $\sim 30$~K to $\sim 50$~K. 
The choice of silicate grain material only mildly affects the temperature in each region with variations well within uncertainties, but it can affect \fac significantly. 
For example, (highly) ferromagnetic-element-rich grain (e.g. containing Fe) refractive indices lead to a higher intrinsic polarisation, and the amount of silicate needed to reproduce the observation is lower than with, e.g., non-Fe-bearing grains.
This is seen by comparing \fac for amorphous MgSiO$_3$ and Mg$_{0.5}$Fe$_{0.5}$SiO$_3$, where the latter yield higher values up to a factor of 2.
However, we can also note that the glassy MgSiO$_3$ yield very similar results as the Mg$_{0.5}$Fe$_{0.5}$SiO$_3$ material. 
Among the five materials presented here, Mg$_{0.7}$SiO$_{2.7}$ yields the lowest \fac values, with only a few percent of carbon material. This suggests that a change in the stoichiometry of the considered material may change its ability to reproduce the observed polarisation, and lead to much higher silicate masses.

In Fig.~\ref{FigFinalMasses}, we present the dust masses derived for $45\degree \leq \theta \leq 90\degree$ using amorphous MgSiO$_3$. Each grain population (silicate vs carbonaceous) follows the same pattern: as $\theta$ increases, the carbon-rich grain mass (empty symbols) increases and the silicate mass (filled symbols) decreases. This is expected since the polarised emission is greatest when the magnetic field is in the plane of the sky (zenith angle 90\degree) and vanishes when it is along the line of sight (zenith angle 0\degree). Thus, a zenith angle of 45\degree\ requires more intrinsic polarised emission to account for the observed polarised intensities. As a consequence, \fac will also increase as a function of $\theta$.

The last two columns in Table~\ref{TabCrabMasses} are upper-limits on the total dust mass in the Crab Nebula, calculated using two different approaches. 
The first method uses the range of carbonaceous and silicate grain masses (columns 3 and 5), and scale the total mass in the three regions to the total area of region 0. Using the total dust map from \citet[][]{DeLooze2019}, we find that the dust mass contained in our three regions is $\sim 34$~per~cent of the total dust mass. We accordingly scale the summed values from the M$_{\rm aC}$ and M$_{\rm Sil}$ to the full map to find the first upper-limit. 
The last column shows the upper-limit on the dust mass using the highest estimated dust surface density between regions~1, 2 and 3, and calculating the corresponding dust mass using the area of region~0. This estimate is naturally higher than the previous method, as it assumes a uniform dust distribution across the remnant. It should therefore be considered as a very conservative upper-limit.

\begin{figure}
    \centering
    \includegraphics[width=0.5\textwidth]{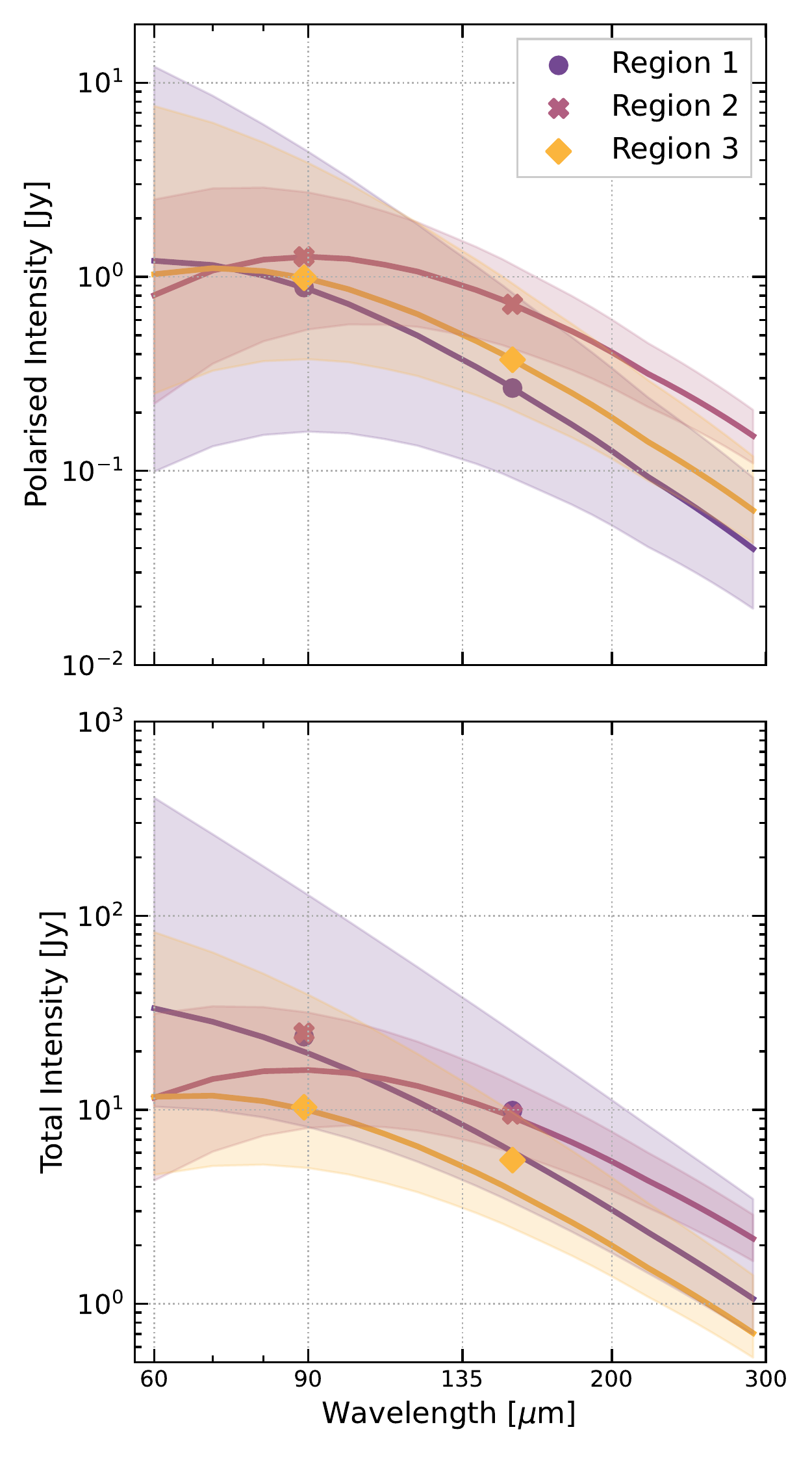}
    \caption{Example SEDs for amorphous MgSiO$_3$ with $\theta=54\degree$. The markers show the observations in each region, and the spectra are computed using the derived \{\tsil, \tac, $\Sigma_{\rm Sil}$, $\Sigma_{\rm aC}$\}, in each region.}
    \label{FigFinalBBodies}
\end{figure}

We use the derived \{\tsil, \tac, $\Sigma_{\rm Sil}$, $\Sigma_{\rm aC}$\} to compute the modelled dust emission spectrum in each case. We find that $\theta=54\degree$ provides a good fit to the data, in the case of amorphous MgSiO$_3$.
In Fig.~\ref{FigFinalBBodies}, we show example SEDs for amorphous MgSiO$_3$ and $\theta = 54\degree$. The shaded regions are computed using the highest and lowest temperatures in each case (from Table~\ref{TabCrabMasses}). The top panel shows the derived silicate dust polarised emission and the bottom panel shows the total dust emission, for all three regions in each panel. 
Despite the fact that we do not fit the data with models, and even with our assumptions, the final results agree fairly closely with the observations.
Additional shorter wavelength data would greatly help constrain the grain temperatures.


\renewcommand{\arraystretch}{1.1}
\begin{table*}
    \centering
    \begin{tabular}{c c c l}
        \textbf{Reference} & \textbf{M$_{\rm dust}$ [\msol]} & \textbf{T$_{\rm dust}$ [K]} & \textbf{Notes} \\
        \hline
        \citet[][]{Gomez2012} & $0.24^{+0.32}_{-0.08}$ & 28 & silicate grains \\
        & $0.11\pm0.01$ & 34 & carbon-rich grains \\
        & 0.14 + 0.08 & - & silicate + carbon-rich grains \\
        \citet[][]{TemimDwek2013} & $0.019^{+0.010}_{-0.003}$ & $56\pm2$ & carbon-rich grains \\
        \citet[][]{OwenBarlow2015} & 0.18--0.27 & - & clumped amorphous carbon-rich grains \\
        & 0.39--0.47 & - & silicate + amorphous carbon-rich grains \\
        \citet[][]{DeLooze2019} & 0.032--0.049 & $41\pm3$ & carbon-rich grains \\
        \citet[][]{Nehme2019} & $0.06\pm0.04$ & - & \\
        \citet[][]{Priestley2020} & 0.05 & - & carbon-rich grains \\
    \end{tabular}
\caption{Compilation of the dust mass estimates in previous works. The temperatures found in this work agree with those in the table. The total dust masses show more variations, and the dust masses derived in this work cover a similar range, as presented in Table~\ref{TabCrabMasses}.}
\end{table*}

\renewcommand{\arraystretch}{1.2}
\begin{table*}
    \centering
    \begin{tabular}{c c c c c c c c}
        & \textbf{\tac [K]} & \textbf{M$_{\rm aC}$ [\msol]} & \textbf{\tsil [K]} & \textbf{M$_{\rm Sil}$ [\msol]} & \textbf{\fac} & M$_{\rm tot}$ [\msol] v1 & M$_{\rm tot}$ [\msol] v2 \\ 
        \hline
        \multicolumn{7}{c}{amorphous MgSiO$_3$ - $\rho = 2.5~$g~cm$^{-3}$}  & \\
        Reg 1 & $67.6\pm38.5$ & 0.0006 -- 0.0009 & $46.9\pm16.1$ & 0.0009 -- 0.0018 & 0.25 -- 0.50 & \hspace{-1em}\rdelim\}{3}{*}[ $<$ 0.040 -- 0.059] & \hspace{-1em}\rdelim\}{3}{*}[ $<$ 0.11] \\
        Reg 2 & $39.0\pm5.2$ & 0.0018 -- 0.0034 & $31.8\pm3.2$ & 0.0065 -- 0.013 & 0.12 -- 0.34 &  & \\
        Reg 3 & $53.2\pm16.1$ & 0.0002 -- 0.0005 & $40.0\pm8.6$ & 0.0013 -- 0.0027 & 0.06 -- 0.25 & &  \\
        \hline
        \multicolumn{7}{c}{glassy MgSiO$_3$ - $\rho = 2.71~$g~cm$^{-3}$} \\
        Reg 1 & $67.1\pm41.1$ & 0.0007 -- 0.0009 & $50.0\pm20.3$ & 0.0004 -- 0.0008 & 0.46 -- 0.70 & \hspace{-1em}\rdelim\}{3}{*}[ $<$ 0.026 -- 0.032] & \hspace{-1em}\rdelim\}{3}{*}[ $<$ 0.061]  \\
        Reg 2 & $38.8\pm4.7$ & 0.0021 -- 0.0036 & $33.0\pm3.4$ & 0.0029 -- 0.0057 & 0.27 -- 0.56 &  \\
        Reg 3 & $52.9\pm16.3$ & 0.0002 -- 0.0005 & $42.1\pm9.7$ & 0.0006 -- 0.0012 & 0.17 -- 0.46 & \\
        \hline
        \multicolumn{7}{c}{amorphous Mg$_{2}$SiO$_{4}$ - $\rho = 3.2~$g~cm$^{-3}$} \\
        Reg 1 & $67.5\pm39.2$ & 0.0006 -- 0.0009 & $46.9\pm16.1$ & 0.0008 -- 0.0016 & 0.29 -- 0.53 & \hspace{-1em}\rdelim\}{3}{*}[ $<$ 0.037 -- 0.054] & \hspace{-1em}\rdelim\}{3}{*}[ $<$ 0.10] \\
        Reg 2 & $38.9\pm5.0$ & 0.0020 -- 0.0035 & $31.8\pm3.3$ & 0.0057 -- 0.0114  & 0.15 -- 0.38 & & \\
        Reg 3 & $53.2\pm16.7$ & 0.0002 -- 0.0005 & $40.0\pm8.9$ & 0.0012 -- 0.0024 & 0.09 -- 0.29 &  & \\
        \hline
        \multicolumn{7}{c}{amorphous Mg$_{0.7}$SiO$_{2.7}$ - $\rho = 2.5~$g~cm$^{-3}$} \\
        Reg 1 & $67.0\pm37.8$ & 0.0006 -- 0.0009 & $45.1\pm14.5$ & 0.0094 -- 0.019 & 0.03 -- 0.08 & \hspace{-1em}\rdelim\}{3}{*}[ $<$ 0.27 -- 0.53] & \hspace{-1em}\rdelim\}{3}{*}[ $<$ 1.0] \\
        Reg 2 & $38.8\pm5.2$ & 0.0012 -- 0.0030 & $31.0\pm3.1$ & 0.066 -- 0.132  & 0.01 -- 0.04 & & \\
        Reg 3 & $52.9\pm16.4$ & 0.0001 -- 0.0004 & $38.7\pm7.9$ & 0.014 -- 0.027 & 0.002 -- 0.03 &  & \\
        \hline
        \multicolumn{7}{c}{glassy Mg$_{0.5}$Fe$_{0.5}$SiO$_3$ - $\rho = 3.2~$g~cm$^{-3}$} \\
        Reg 1 & $67.1\pm37.4$ & 0.0007 -- 0.0009 & $50.9\pm19.8$ & 0.0004 -- 0.0008 & 0.44 -- 0.68 & \hspace{-1em}\rdelim\}{3}{*}[ $<$ 0.027 -- 0.034] & \hspace{-1em}\rdelim\}{3}{*}[ $<$ 0.065] \\
        Reg 2 & $38.8\pm4.9$ & 0.0022 -- 0.0036 & $33.4\pm3.6$ & 0.0031 -- 0.0062 & 0.26 -- 0.54 &  & \\ 
        Reg 3 & $52.9\pm16.4$ & 0.0003 -- 0.0005 & $42.8\pm10.2$ & 0.0006 -- 0.0013 & 0.17 -- 0.44 &  & \\
        \hline
    \end{tabular}
    \caption{Compilation of the dust mass estimates in this study. We assume a unique carbonaceous material: optECs with $E_g=0.1~$eV \citep[][]{Jones2012a, Jones2012b, Jones2012c}.
    The large errors on the temperatures are due to the large errors in observed polarised intensity and polarisation fraction. The grain masses and fractions of carbonaceous are given as the range of masses for $45\degree \leq \theta \leq 90\degree$. Note that M$_{\rm aC}$ and M$_{\rm Sil}$ move in opposite direction: the low value of M$_{\rm aC}$ is paired with the high value of M$_{\rm Sil}$ (as shown in Fig.~\ref{FigFinalMasses}).
    The last two columns are upper-limits on the total dust mass, using two different approaches (see Section~\ref{SecPolResults}).}
    \label{TabCrabMasses}
\end{table*}

\section{Discussion}
\label{SecDiscussion}
\subsection{Previous work}
Previous works have derived dust masses in the Crab Nebula with other methods. For instance, fitting the total SED from near-IR to radio of the Crab Nebula, \citet[][]{Gomez2012} found $0.14$ plus $0.08$~\msol of carbonaceous and silicate dust, respectively, while \citet[][]{DeLooze2019} found between 0.032 and 0.049~\msol of dust, using resolved (pixel-by-pixel) SEDs. Using radiative transfer, \citet[][]{TemimDwek2013} found a total dust mass of 0.019~\msol, \citet[][]{OwenBarlow2015} found  $0.11$--$0.13$ plus $0.39$--$0.47$~\msol of carbonaceous and silicate, respectively, and \citet[][]{Priestley2020} found a mass of carbonaceous grains of $\sim 0.05$~\msol.
The different approaches taken in these works all yield different values, sometimes significantly, exemplifying the difficulty to constraint dust masses. 
In \citet[][]{DeLooze2019}, the authors performed a pixel-by-pixel SED fit to the Crab at 36$''$ resolution using carbonaceous grains only (the same ones used in this work). They found total dust masses of 0.0056~\msol in region~1, 0.0051~\msol in region~2, and 0.0028~\msol in region~3. The closest masses from Table~\ref{TabCrabMasses} are those using ${\rm Mg_{0.7}SiO_{2.7}}$ as silicate material, but for which the carbon-rich fractions are particularly low. Also note that \citet[][]{DeLooze2019} discredited this material as it yielded unrealistic values with their approach (on integrated scales).
Overall, the masses found here are lower than those from their paper. This is likely due to the higher dust temperatures found in this work, and the properties of the grains (e.g., enstatite being less emissive than the carbon-rich grains at these wavelengths).

\subsection{On the variations of \textit{p}}
In Fig.~\ref{FigPlotpValues} and Table~\ref{TabMeanPol} we report the values of the polarisation fraction in each region.
The dust-only polarisation fraction is higher at 89~$\mu$m than 154~$\mu$m in regions~1 and 3 (3.7 and 9.6 per cent, 2.7 and 6.8, respectively), but lower in region~2 (5.1 and 7.6 per cent, see Table~\ref{TabMeanPol}).
It is, however, relatively difficult to conclude on the significance of these differences given the large errors on $p$.

Differences in polarisation degree can come from the grain characteristics like their size, axis ratios, or composition, but also environmental properties like the radiation field, or magnetic field geometry.
The work of \citet[][and references therein]{Matsumura2011} suggests that higher polarisation can be due to stronger radiation heating the dust grains.
Three dimensional reconstruction of the Crab \citep[e.g.,][]{Martin2021} could help localise the filaments with respect to the heating source, helping to situate each region and interpret the variations seen in $p$.

\subsection{On the dust composition constraints and \costuum inputs}
Some of the assumptions described in Section~\ref{SecConstraints} can be changed as inputs of the \costuum tool.

\textit{Dust analogues --}
Our method can be applied to any dust-like material with \costuum, so long that refractive indices and volumetric density are available. Our results show that a change in the dust analogue mostly affects the value of \fac (Table~\ref{TabCrabMasses}).
In \citet[][]{Arendt2014}, the authors found that glassy ${\rm MgFeSiO_4}$ with $\rho=3.71$g~cm$^{-3}$ \citep[][]{Jaeger1994, Dorschner1995} or ${\rm Mg_{0.7}SiO_{2.7}}$ \citep[]{Jager2003}\footnote{we assume a volume density of 2.5~g~cm$^{-3}$.} provide good fits to several spectra of the Cassiopeia~A supernova remnant.
Here, using ${\rm MgFeSiO_4}$ reaches similar \fac as using glassy ${\rm MgSiO_3}$ or ${\rm Mg_{0.5}Fe_{0.5}SiO_3}$, while amorphous ${\rm Mg_{0.7}SiO_{2.7}}$ dramatically reduces \fac (Table~\ref{TabCrabMasses}).
Work by \citet[][]{Demyk2017a, Demyk2017b} measured mass absorption coefficients of a variety of glassy (iron rich-)silicates lower than, e.g., the amorphous MgSiO$_3$ material used in Table~\ref{TabCrabMasses}.
Similarly, note that \citet[][]{Jones2017} use silicate grain densities reduced by 14--30~per~cent when considering nano-particles, as opposed to bulk material properties found in the JENA database. In these cases, we would find lower dust masses.
Additionally, note that previous works have discredited the presence of glassy material in SNRs \citep[e.g.][]{Rho2008, Arendt2014, Rho2018}. The absorption cross-section shows features too sharp to reproduce the observations of SNRs.

In Section~\ref{SecPolResults}, we showed that changing the stoichiometry of the grain species affects the results. This suggests that it is possible to explain the data with a single dust component, i.e., grains giving polarised emission are giving all of the total emission. This follows the hypothesis of the ``astrodust'' model by \citet[][]{DraineHensley2021Astrodust}. 
By adjusting extinction and emission measurements, with constraints from depletion measurements, they derive a unique material able to reproduce observations. The astrodust material is partially composed of amorphous silicate but also contains carbon-rich material. 
We performed similar investigation using their astrodust model with porosity $\mathcal{P}=0.20$, Fe-inclusion $f_{\rm Fe} = 0.0$, and axis ratio $b/a=1.4$. Because of the different approach (a unique dust component), we cannot follow the same method described in Section~\ref{SecMethod}, and introduce instead a fraction of aligned grains, $f_{\rm align}$, responsible for the polarised signal. More details are given in Appendix~\ref{AppAstrodust}. With the astrodust component, we find dust temperatures $T_{\rm dust} = 37.2, 39.2, 32.4$~K in regions~1, 2, and 3, and corresponding dust masses of $M_{\rm dust} = 6.9\times10^{-3}, 5.3\times10^{-3}, 4.1\times10^{-3}$~\msol, with $f_{\rm align} = 0.13, 0.26, 0.34$. Fig.~\ref{AppFigAstrodust} demonstrates that the astrodust model is able to reproduce the observations as well as the two-component approach: after constraining the temperature of the grains using total intensity, the derived emission models fit the data very well. The best fit being that in total or polarised intensity depends on which constrain to use to solve for the astrodust grain temperatures. We do caution, however, that the number of data points is limited relative to the number model parameters, preventing a more stringent test of the one- vs two-component paradigms.

\textit{Grain size distribution~--~}
The nature and detailed parameters of grain size distributions in SNe and SNRs are subject to substantial variations. Different works found that the overall size distribution can be a power-law or log-normal \citep[e.g.,][]{TemimDwek2013, OwenBarlow2015, Bocchio2016} or that individual species may be modelled by a different law than the overall size distribution \citep[e.g.,][]{Nozawa2003, Priestley2020}. In Section~\ref{SecConstraints}, we present results for a power-law size distribution with index of 3.5. Here we briefly compare the results for MgSiO$_3$ using different size distribution assumptions.

If we use a shallower index of 2.5, the fraction of carbon-rich grains, \fac, increases in all regions, by 3--5~per~cent only. The carbonaceous and silicate grain temperatures decrease by a few K, not significant considering the uncertainties. 
Using a power-law index of 4.5 has the opposite effect, and \fac decreases by $\sim 10$~per~cent. The slope of the power-law size-distribution does not affect the total dust mass upper-limit outside of the uncertainty.

Future work should investigate how the minimum and maximum grain radius (here 0.1 and 5~$\mu$m) affect these results. It will be interesting to include non-polarising small silicate grains, as their contribution to the total intensity, but not to the polarised intensity will likely decrease the observed fraction of carbonaceous grains.

\textit{Shape distribution --}
In \costuum, we assumed the CDE2 shape distribution from \citet[][median axis ratio of 2.73]{Ossenkopf1992}, though the most extreme shapes are not used.
By sampling (even) less elongated grains, the polarisation from silicate grains decreases and the overall \fac values decreases only by a few ~per~cent for MgSiO$_3$.
We also ran tests assuming a single shape distribution, with axis ratios $d$ of 0.5 and 2.0, which effectively change the grain from prolate to oblate. We find that assuming $d=0.5$ leads to slightly lower \fac by $\sim 5$~per~cent, while $d=2.0$ has the opposite effect, leading to higher \fac.
This goes in the same direction as the work of \citet[][]{Kirchschlager2019Pint} who found differences in polarisation between oblate and prolate grains (porous grains with $d=1/1.5$ and $d=1.5$, respectively), with slightly higher polarisation for oblate grains.
We also refer the reader to \citet[][]{Vandenbroucke2020} for an extensive review of how $Q_{\rm abs}$ varies with different inputs in \costuum.

\section{Conclusions}
In this paper, we present new SOFIA/HAWC+ polarimetric observations of the Crab Nebula in bands~C (89~$\mu$m, FWHM~$\sim 7.8''$, and D~(154~$\mu$m, FWHM~$\sim 13.6''$) (Fig.~\ref{FigDataStokes}).
We report detected polarised emission from this supernova remnant, when integrating the signal in three dusty filaments, in order to boost the S/N of individual pixels (Fig.~\ref{FigPolRebin}. This constitutes the second SNR, after Cassiopeia~A, for which polarisation detection is confirmed.
This polarised emission implies the presence of grains large enough to be efficiently aligned by the local magnetic field, with radii $a \gtrsim 0.05-0.1~\mu$m assuming efficiencies comparable to the diffuse ISM, but bringing more constraints to that value proves difficult.

After convolution to the SPIRE~500 resolution ($36''$), we remove the synchrotron contribution (Fig.~\ref{FigfSync}), and obtain synchrotron-free Stokes maps and polarisation vectors (\ref{FigFinalPol}). 
We find an average polarisation of 2.7 and 3.7~per~cent in the whole maps, and averages from 2.7 to 9.6~per~cent in three individual regions exhibiting high fluxes in the far-IR bands (Fig.~\ref{FigPlotpValues} and Table~\ref{TabMeanPol}), identified as dusty filaments.

Using total flux and average polarisation in each region, we use the simulation tool \costuum to derive dust temperatures and masses in three regions of interest (Table~\ref{TabCrabMasses}). We use a unique carbonaceous population and five silicate materials. We find carbonaceous grain masses ranging from $10^{-4}$ to $\sim 7 \times 10^{-2}$~\msol, and silicate grain masses spanning $10^{-4}$ to $10^{-1}$~\msol. These masses lead to fraction of carbon-rich grains as low as 1~per~cent up to 80~per~cent. Dust temperatures prove rather insensitive to the chosen dust material, and span ${\rm \sim 40~K \leq T_{aC} \leq \sim 70~K}$ and ${\rm \sim 30~K \leq T_{Sil} \leq \sim 50~K}$.

The work presented here makes use of the polarimetry of only two photometric bands, sampling the IR peak. Precisely estimating dust characteristics remains very difficult at these S/N, which could be increased with additional integration time. The decommissioning of the SOFIA aircraft and the lack of a next generation of more sensitive IR polarimeters limit further investigation.

\section*{Acknowledgements}
The authors are thankful to the referee for their careful reading of the initial and updated drafts, and their suggestions which helped improved the paper.
The authors thank Anthony Jones for sharing the optECs properties of carbonaceous grains, and his comments on the first draft of the paper. We thank Eli Dwek for sharing optical grain properties of iron-type grains.
JC and IDL acknowledge support from ERC starting grant \#851622 DustOrigin.
HLG acknowledges support from ERC consolidator grant CosmicDust. MR acknowledges support from project PID2020-114414GB-100, financed by MCIN/AEI/10.13039/501100011033. KP is a Royal Society University Research Fellow, supported by grant no. URF\textbackslash R1\textbackslash211322.
This research is based on observations made with the NASA/DLR Stratospheric Observatory for Infrared Astronomy (SOFIA). SOFIA is jointly operated by the Universities Space Research Association, Inc. (USRA), under NASA contract NNA17BF53C, and the Deutsches SOFIA Institut (DSI) under DLR contract 50 OK 0901 to the University of Stuttgart.
This research made use of \texttt{matplotlib}, a Python library for publication quality graphics \citep{Hunter:2007};
Astropy, a community-developed core Python package for Astronomy \citep{2018AJ....156..123A, 2013A&A...558A..33A};
NumPy \citep{van2011numpy};
SciPy \citep{Virtanen_2020};
APLpy, an open-source plotting package for Python \citep[][]{aplpy12, aplpy19}.
\section*{Data Availability}
The SOFIA/HAWC+ observations presented here are available through the SOFIA Science Archive on the IRSA website.
The synchrotron-subtracted maps are available upon request to the first or second author.

\bibliographystyle{mnras}
\bibliography{example} 

\appendix
\section{Multi-studies map comparison}
\label{AppOverplots}
Here, we show: (i) the SOFIA/HAWC+ bands C and D Stokes I maps after subtraction of the synchrotron emission, and the associated polarisation vectors (top row); (ii) the NIKA~15-GHz Stokes I map, and the associated polarisation vectors (bottom left); (iii) the total dust mass map from \citet[][]{DeLooze2019}. The contours show the threshold $M_{\rm dust} = 10^{-4}$~\msol. The ellipses are the three regions of interest described in Section~\ref{SecResults}.

\begin{figure*}
    \centering
    \includegraphics[width=\textwidth, clip, trim={1cm 0 1.0cm 0}]{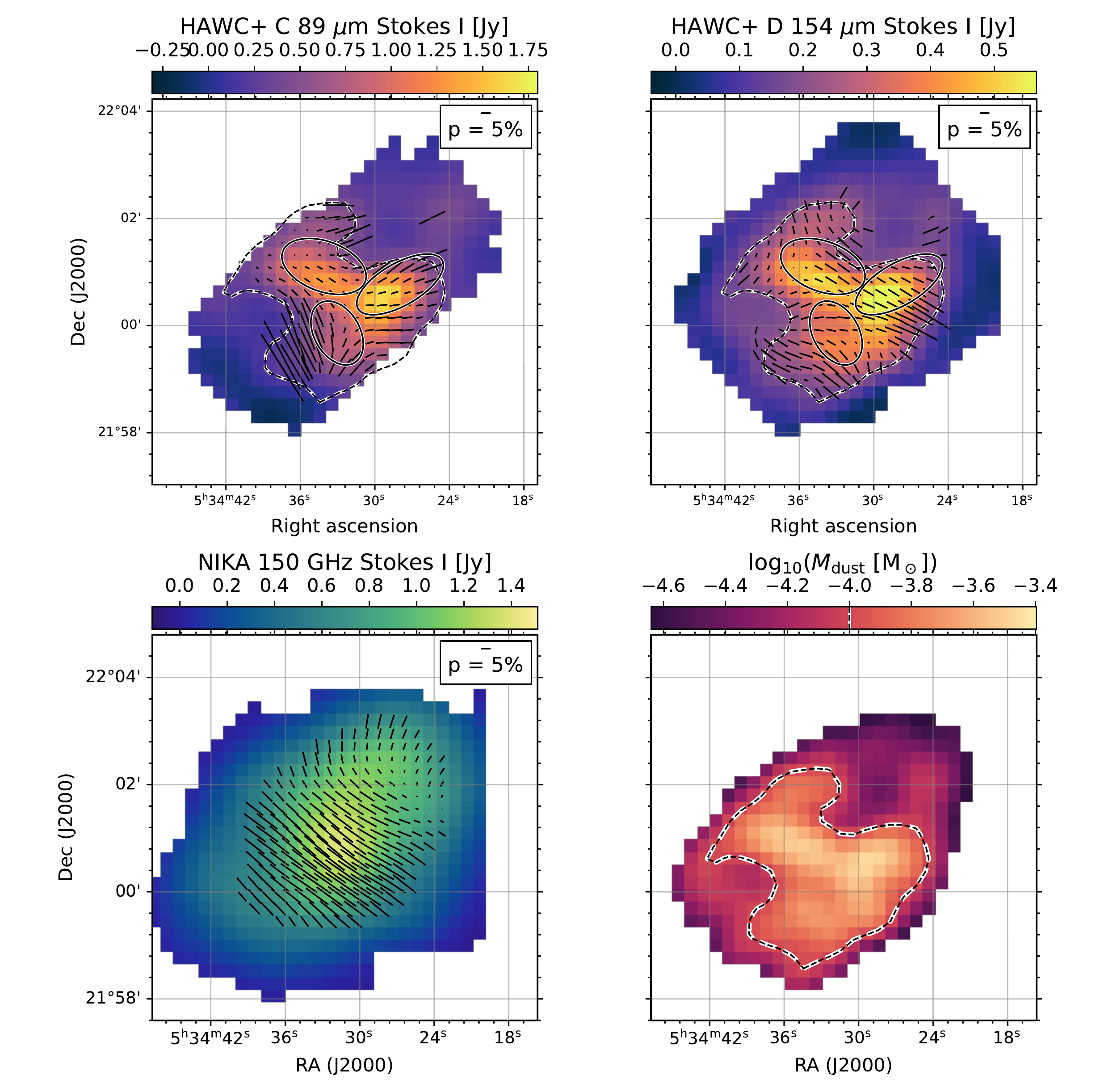}
    \caption{\textit{Top row:} Synchrotron-subtracted Stokes I parameter maps for band~C (left) and band~D (right), with dust-only polarisation vectors. \textit{Bottom left:} NIKA~150~GHz Stokes I parameter map with polarisation vectors. \textit{Bottom right:} Total dust mass map from \citet[][]{DeLooze2019} where the contours mark the M$_{\rm dust} = 0.0001$~\msol per pixel. All maps are at the SPIRE~500 36$''$ resolution.}
    \label{AppPanels}
\end{figure*}

\section{Using the ``astrodust'' component}
\label{AppAstrodust}
Combining carbonaceous and silicate-rich grains, \citet[][]{DraineHensley2021Astrodust} created a unique component, ``astrodust'' that is able to reproduce Milky Way observations from UV to microwave wavelengths. Extinction and emission measurements \citep[][]{HensleyDraine2020, HensleyDraine2021}, coupled with depletion constraints, are used to find a unique dielectric function that reproduce all observables, including polarisation constraints. In this model, a single astrodust grain contains distinct domains of both amorphous silicate and carbonaceous materials. We refer the reader to \citet[][]{DraineHensley2021Astrodust} for more details, e.g., on grain shapes, or porosity\footnote{The data is available at \url{https://doi.org/10.34770/9ypp-dv78}.}.
Here, we present a brief test using this unique component to reproduce polarisation observation in the Crab Nebula, instead of the dual grain population carbonaceous + silicate grains. 
Because the main assumption in Section~\ref{SecMethod} relies on having two grain populations with only one population responsible for the polarised emission (to be able to cancel out most terms), we cannot use the same solving methodology when using a single component. Instead, we introduce a fraction of aligned grains, $f_{\rm align}$. With that assumption, we proceed as follow, in each region:
\begin{enumerate}
    \item we solve the dust temperature $T_{\rm dust}$ using the same approach as before, i.e., with the ratio of the total intensities at 89 and 154~$\mu$m;
    \item we derive the corresponding dust mass to reproduce the observed total intensity;
    \item we solve $f_{\rm align}$ using the observed polarisation and the astrodust mass absorption coefficients.
\end{enumerate}

\begin{figure*}
    \centering
    \includegraphics[width=\textwidth]{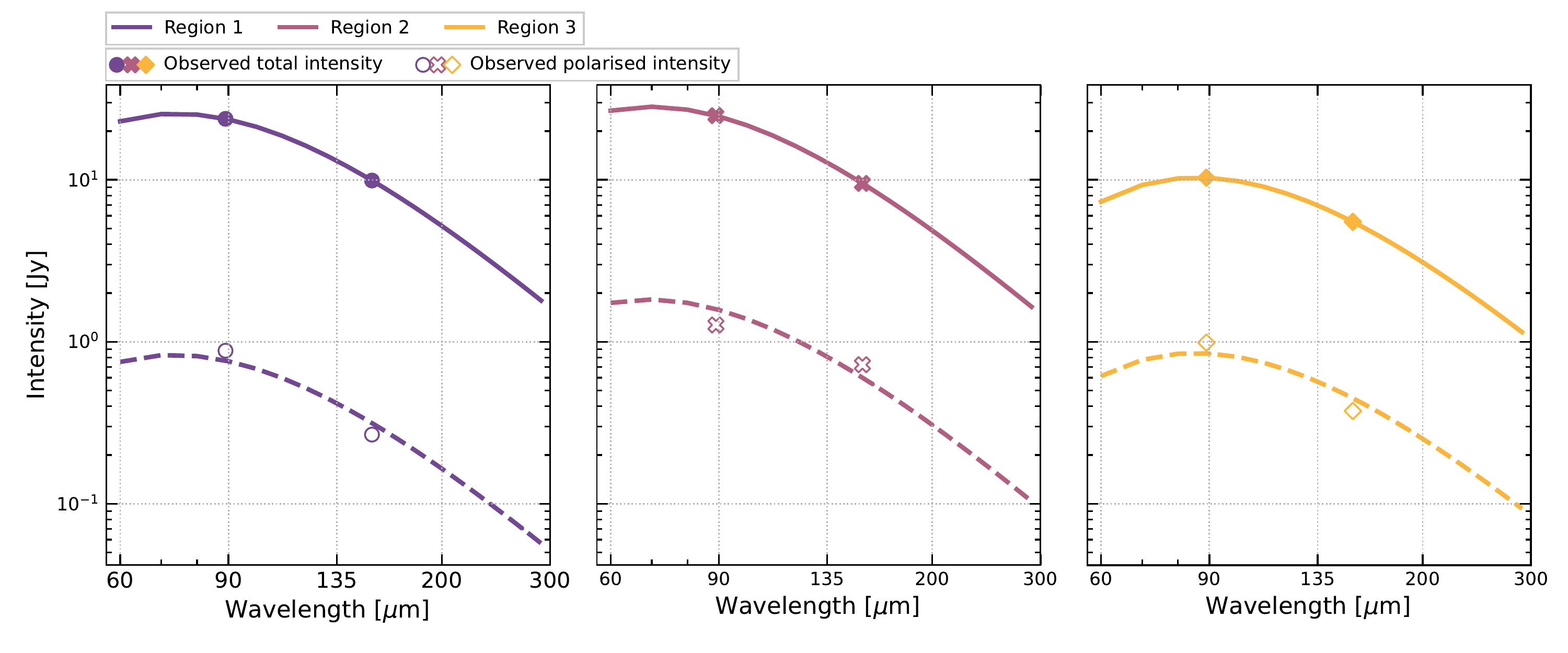}
    \caption{Example SEDs for the astrodust component \citep[][]{DraineHensley2021Astrodust}. The markers show the observations in polarised intensity (empty symbols) and total intensity (filled symbol). The spectra are computed using the dust temperature and mass found using the total intensity observations, and fraction of aligned grains using the observed polarisation fractions. The solid lines show the total intensity, and the dashed lines the polarised intensity, after applying the $f_{\rm align}$ correction.}
    \label{AppFigAstrodust}
\end{figure*}

We find dust temperatures $T_{\rm dust} = 37.2, 39.2, 32.4$~K, dust masses $M_{\rm dust} = 6.9\times10^{-3}, 5.3\times10^{-3}, 4.1\times10^{-3}$~\msol, and fraction of aligned grains $f_{\rm align} = 0.13, 0.26, 0.34$, in regions~1, 2, and 3, respectively.
The temperatures found using astrodust are close to those of the silicate grains in Section~\ref{SecResults}, and show less variations between each region than when using the two-population approach (where they can vary by $\sim 15$~K). In Fig.~\ref{AppFigAstrodust} we show the derived dust SEDs using the temperatures and masses in each region, for the total (solid lines, filled symbols) and the polarised intensities (dashed lines, empty symbols). Note that the temperatures and masses are derived using the total intensities, and used to model the polarised intensities, applying $f_{\rm align}$. 
On the basis of this figure, it appears that the astrodust hypothesis is also able to explain the polarised emission in the Crab Nebula. Caution should be applied considering these results, and more work and observations are needed to properly constrain the dust properties using astrodust.

\bsp	
\label{lastpage}
\end{document}